\documentclass[nofootinbib,aps,a4paper,11pt,letterpaper,superscriptaddress,onecolumn,showkeys,times,eqsecnum]{revtex4-2}
\usepackage{amsmath}
\usepackage{amsfonts}
\usepackage{tabularx}
\usepackage[utf8]{inputenc}
\usepackage{amsmath}
\usepackage{colortbl}
\usepackage{array}       
\usepackage{tabularx}     
\usepackage{lipsum} 
\usepackage{array}
\usepackage{multirow}
\usepackage{makecell}
\usepackage{booktabs}
\usepackage{graphicx}
\usepackage{multirow}
\usepackage{color}
\usepackage{braket}
\usepackage{dcolumn}
\usepackage{siunitx}
\usepackage{bm,url}
\usepackage{colortbl}
\usepackage[linktocpage]{hyperref}
\usepackage{subfigure}
\usepackage{amsfonts}
\usepackage[left=2.5cm,right=2.5cm,top=2.5cm,bottom=2.5cm]{geometry}
\usepackage[usenames,dvipsnames,svgnames]{xcolor}  
\usepackage{hyperref}   
\definecolor{oxfordblue}{rgb}{0.0, 0.13, 0.28}
\definecolor{burgundy}{rgb}{0.5, 0.0, 0.13}
\definecolor{darkolivegreen}{rgb}{0.33, 0.42, 0.18}
\definecolor{darkblue}{rgb}{0,0,0.5}
\definecolor{richcarmine}{rgb}{0.84, 0.0, 0.25}
\definecolor{darkblue}{rgb}{0,0,0.5}
\definecolor{bluer}{rgb}{0.00,0.50,0.75}{}
\hypersetup{colorlinks=true, citecolor=red, linkcolor=blue,
	urlcolor = magenta, filecolor=magenta}

\begin{document}

\title{Pathology of the Unified Dark Sectors in Modified General Relativity }
\author{\textbf{Mohsen Khodadi}}
\email{m.khodadi@du.ac.ir}
\affiliation{School of Physics, Institute for Research in Fundamental Sciences (IPM),\\ P.~O.~Box 19395-5531, Tehran, Iran}
\affiliation{School of Physics, Damghan University, Damghan 3671641167, Iran}
\affiliation{Center for Theoretical Physics, Khazar University, 41 Mehseti Str., AZ1096 Baku, Azerbaijan}

\begin{abstract}
This paper presents a comprehensive stability analysis of the black hole solution within Modified General Relativity (MGR), a theory proposing a unified geometric description of dark matter (DM) and dark energy (DE). A rigorous gauge-invariant formalism is employed to analyze gravitational perturbations of the extended Schwarzschild metric. The central finding is a critical pathology within the polar perturbation sector, where metric fluctuations couple to the theory's fundamental line element field. This coupling is governed by a factor that, while well-behaved at the horizon, diverges powerfully in the far-field limit as a direct consequence of the theory's non-asymptotically flat nature. This indicates a strong infrared instability that overwhelms perturbations at large distances. In stark contrast, the axial perturbation sector is found to be completely stable. This dichotomy proves that the instability is not inherent to the background metric but is specifically generated by the novel coupling mechanism encoding MGR's unified dark sectors. The results reveal a fundamental strong-coupling problem within the MGR framework, challenging its physical viability as an alternative to Einstein's General Relativity (EGR).	
	
\vspace{0.5cm}
\textbf {Keywords:} Modified General Relativity, Dark matter, Dark energy, Black hole stability, Gravitational perturbations, Strong coupling problem
\end{abstract}
\maketitle
\section{Introduction}
\label{sec:intro}

The composition of our universe presents a profound mystery, one that has fundamentally reshaped the pillars of modern cosmology and particle physics. For centuries, humanity's understanding of the cosmos was limited to the material that emits light: stars, galaxies, and gas. However, a series of groundbreaking observations throughout the 20th and 21st centuries have revealed that the luminous matter we can see constitutes less than $5\%$ of the total mass-energy content of the universe. The remaining $~95\%$ is composed of two enigmatic components: dark matter (DM) and dark energy (DE), e.g., see review papers \cite{Frieman:2008sn,Bertone:2016nfn,Oks:2021hef}. These entities, while invisible and detectable only through their gravitational influence, are the dominant architects of the universe's structure and its ultimate fate.

The prevailing hypothesis for DM is that it is a yet-undetected particle. However, a parallel and compelling line of inquiry asks: What if DM doesn't exist? What if, instead, our theory of gravity—Einstein's General Relativity (EGR)—is incomplete, and we are misinterpreting its effects on cosmic scales as evidence for an invisible substance?\cite{Clifton:2011jh} This is the core premise of Modified Newtonian Dynamics (MOND) and its relativistic extensions, collectively known as Modified Gravity (MG) theories. They propose that the laws of gravity change at very low accelerations, precisely where the DM discrepancy appears. The foundational idea was proposed by Milgrom \cite{Milgrom:1983ca} (see also \cite{Bekenstein:1984tv}).

Among recent proposals, a theory named Modified General Relativity (MGR) has been advanced by Gray Nash \cite{Nash:2019rvd}. This theory falls within the MG paradigm and offers a unified geometric description of DM and DE. The framework of MGR is built upon the introduction of a new connection-independent tensor, \(\Phi_{\alpha\beta}\), which is argued to represent the energy-momentum of the gravitational field itself. This tensor is constructed from a fundamental line element field \((\mathbf{X}, -\mathbf{X})\), a feature inherent to any Lorentzian spacetime. A key success claimed for MGR is the natural derivation of a force law that reproduces the empirical Tully-Fisher relation, a cornerstone of MOND phenomenology \footnote{Although MGR phenomenologically on galactic scales is alignment with MOND, conceptually it contrasts.}. Furthermore, the trace of this new tensor, \(\Phi\), dynamically replaces the cosmological constant \(\Lambda\), with the condition \(\Phi > -2\Phi_{00}\) describing repulsive dark energy. Consequently, MGR posits a single geometric framework capable of simultaneously describing DM and DE \cite{Nash:2019rvd}. Other success of MGR is that it fix two flaws in EGR i.e., non-localization of gravitational energy, and ontological inconsistency \footnote{Concerning the former, in EGR unlike electromagnetism, one can know the total energy of a system, but cannot meaningfully define how that energy is distributed from point to point in space. According to Gary Nash's claims, this solves this by finally providing a proper tensor for gravitational energy. The latter means that the Einstein equation has a geometric object ($G_{\mu\nu}$) on one side, describing the curvature of spacetime, and a non-geometric matter source ($T_{\mu\nu}$) on the other. Nash quotes Einstein's own dissatisfaction \cite{Ei}, comparing it to a building with one wing made of "fine marble" and the other of "low-grade wood." The left-hand side describes the self-interaction of the dynamic metric, while the right-hand side has nothing that represents this dynamism.}. One of the most interesting features of MGR is that it mathematically connects geometric and particle perspectives on DM, showing they are not mutually exclusive \cite{Nash:2019hru,Nash:2023zza,Nash:2025xhe}. Specifically, it explains rotation curves through a dual geometric and particle view of DM.

Given these ambitious claims, a rigorous assessment of the theory's physical viability is essential. A critical test for any gravitational theory is the stability of its black hole solutions. The theoretical and phenomenological implications of the MGR black hole solution necessitate a detailed stability analysis against spacetime perturbations.

This manuscript presents a comprehensive linear stability analysis of the spherically symmetric black hole solution in MGR. The analysis is structured as follows: Sec. \ref{Re} provides a review of the extended Schwarzschild black hole solution within MGR. Sec. \ref{polar} details the perturbation formalism, employing  the Regge-Wheeler gauge, and gauge-invariant approaches for general spherically symmetric spacetimes \cite {Kodama:2003jz} to derive a master equation for polar (even-parity) perturbations. This derivation explicitly includes the perturbations of the new fundamental field in MGR, leading to a sourced wave equation. Subsequently, in Sec. \ref{f}, we examine the asymptotic behavior of the perturbations in the line element field, considering both the near-horizon and far-field limits. Section \ref{sta} examines the asymptotic structure of the spacetime, derives a conserved energy flux, and provides a numerical demonstration of the pathological growth driving the instability. 
For completeness and to isolate the origin of any instability, the analysis is extended to axial (odd-parity) perturbations in Sec. \ref{axi}, where a homogeneous master equation is derived. Finally, Sec. \ref{con} discusses the results, contextualizes the findings within the broader landscape of modified gravity theories—including the relation to strong-coupling problems and screening mechanisms \cite{Vainshtein:1972sx}—and presents the concluding remarks on the viability of MGR.

\section{Review of Black Hole Solution in MGR}\label{Re}

This review is based on the Ref. \cite{Nash:2019rvd} which extends GR by introducing a connection-independent tensor \(\Phi_{\alpha\beta}\) to represent gravitational energy-momentum. The key steps involve solving the field equations in a spherically symmetric, static spacetime to obtain an extended Schwarzschild solution.

The modified Einstein equation in MGR is
\begin{equation}
G_{\alpha\beta} + \Phi_{\alpha\beta} = \frac{8\pi G}{c^4} \tilde{T}_{\alpha\beta},
\end{equation}
where \(\Phi_{\alpha\beta}\) is constructed from the Lie derivative of the metric and the line element field \((X, -X)\):
\begin{equation} \label{dif}
\Phi_{\alpha\beta} = \frac{1}{2} \pounds_X g_{\alpha\beta} + \pounds_X (u_\alpha u_\beta),
\end{equation} where by expanding the Lie derivatives, it reads off
\begin{equation}\label{dif1}
\Phi_{\mu\nu} = \frac{1}{2} \left( \nabla_\mu X_\nu + \nabla_\nu X_\mu \right) + u^\lambda \left( u_\mu \nabla_\nu X_\lambda + u_\nu \nabla_\mu X_\lambda \right).
\end{equation}
The line element field \(X^\alpha\) is collinear with the unit timelike vector \(u^\alpha=(u^0,0,0,0)\):
\begin{equation}
X^\alpha = f u^\alpha, \quad u^\alpha u_\alpha = -1.
\end{equation}
where the function $f$ represents the magnitude of the line element field $X^{\alpha}$ i.e., $X^{\alpha}X_{\alpha}=-f^2$. Note that the variation of MGR action result in to following constraint
\begin{equation}\label{cons}
\nabla_\alpha (u^\alpha u^\beta)=0, ~~~~ \mbox{or}~~~\nabla_\alpha u^\alpha=0.
\end{equation}
along with 
\begin{equation}\label{cons1}
\Phi=-u^\alpha\partial_\alpha f
\end{equation} which $\Phi$ is trace of \(\Phi_{\alpha\beta}\).

For a vacuum (\(\tilde{T}_{\alpha\beta} = 0\)), the field equations reduce to:
\begin{equation}
G_{\alpha\beta} + \Phi_{\alpha\beta} = 0.
\end{equation}
We assume a static, spherically symmetric metric in Schwarzschild coordinates:
\begin{equation} \label{ss}
ds^2 = -e^{\nu(r)} c^2 dt^2 + e^{\lambda(r)} dr^2 + r^2 (d\theta^2 + \sin^2\theta d\phi^2). 
\end{equation}
For simplicity, we set \(\nu(r) = -\lambda(r)\) due to spherical symmetry, as in GR. In the static case, the field components are assumed to be
\begin{equation}
X_0 = f u_0, \quad X_1 = f u_1, \quad X_2 = X_3 = 0.
\end{equation}
From the normalization condition \(u^\alpha u_\alpha = -1\), we have:
\begin{equation}
u_0 u^0 + u_1 u^1 = -1.
\end{equation}
The non-zero components of \(\Phi_{\alpha\beta}\) are:
\begin{align}
\Phi_{00} = \frac{\mu}{2} e^{-2\lambda} \lambda' X_1, \quad \Phi_{11} = (1 + 2u_1 u^1) \left( X_1' - \frac{\lambda'}{2} X_1 \right),\\
\Phi_{22} = (1 + 2u_2 u^2) \left( \partial_2 X_2 + r e^{-\lambda} X_1 \right), \quad \Phi_{33} = \sin^2\theta \Phi_{22},
\end{align}
where \(\mu = 1 + 2u_0 u^0\). 

The Einstein tensor components for the metric are:
\begin{align}
G_{00} =& \frac{e^{-2\lambda}}{r^2} (r \lambda' - 1) + \frac{e^{-\lambda}}{r^2},\\
G_{11} =& \frac{1}{r^2} (1 - r \lambda') - \frac{e^{\lambda}}{r^2},\\
G_{22} = &\frac{r^2 e^{-\lambda}}{2} \left( -\lambda'' + \lambda'^2 - \frac{2 \lambda'}{r} \right),\\
G_{33} = &\sin^2\theta \, G_{22}.
\end{align}

For \((\alpha, \beta) = (0,0)\) and \((1,1)\):
\begin{equation}
\mu \frac{\lambda'}{2} X_1 + (X_1' - \frac{\lambda'}{2} X_1)(1 + 2u_1 u^1) = 0.
\end{equation}
For \((\alpha, \beta) = (2,2)\):
\begin{equation}
-\lambda'' + \lambda'^2 - \frac{2 \lambda'}{r} + \frac{2 e^{\lambda}}{r^2} \left( \partial_2 X_2 + r e^{-\lambda} X_1 \right)(1 + 2u_2 u^2) = 0.
\end{equation}
By assuming a power series expansion for \(X_1\), and \(X_2\)
\begin{align}
X_1 =& e^{\lambda} \left( a_0 + \frac{a_1}{r} + \frac{a_2}{r^2} \right),\\
X_2 = &\left( b_0 + \frac{b_1}{r} \right) \tan\theta,
\end{align}
after substitution into the field equations and simplification, the following equation is obtained
\begin{equation}
-\lambda'' + \lambda'^2 - \frac{2 \lambda'}{r} + \frac{2 e^{\lambda}}{r^2} (a_0 r - b) = 0,
\end{equation}
where \(b = -(a_1 + b_0)\). Note that \(a_{0,1,2}\) and \(b_{0,1}\) are constants so that the former ensures the gravitational energy density behaves as \(\sim 1/r^4\) at large distances. Before solving the above equation, it must be mention that the field equations \(G_{\alpha\beta} + \Phi_{\alpha\beta} = 0\) yield the the trace of gravitational energy-momentum tensor as follows
\begin{equation}\label{R}
	\Phi=g^{\alpha\beta}\Phi_{\alpha\beta}=e^{-\lambda}\bigg(\lambda''-\lambda'^2+\frac{4\lambda'}{r}-\frac{2}{r^2}\bigg)+\frac{2}{r^2},
\end{equation} where is equals Ricci scalar $R$.

The exact solution takes the following extended form
\begin{equation}\label{metric}
e^{-\lambda} =e^{\nu}=1-\frac{2GM}{c^2r} + 2b \ln r-a_0r ,
\end{equation}
This solution enriched with two free parameters due to adding $\Phi_{\alpha\beta}$ in standard GR action with the following observational functions

1. \(a_0\): DE parameter, scaling with mass as:
\begin{equation} \label{a}
|a_0| = \frac{\zeta M l_p}{M_\odot}, \quad \zeta = 3.55 \times 10^{-6} \, \text{m}^{-2}.
\end{equation}
It potentially has two sings: $a_0>0$, describes the repulsive DE
energy force in the current epoch, while $a_0<0$ address a strongly
past deceleration epoch, confirmed by Riess et al., \cite{SupernovaSearchTeam:2004lze}. 
 
2. \(b\): DM parameter (dimensionless):
\begin{equation}\label{b}
b = -\sqrt{\frac{|a_0| |2GM/c^2|}{2}}~~\Longrightarrow b=-\sqrt{|a_0|M} ~~~(\mbox{natural unit $G=1=c$}).
\end{equation}
Equation (\ref{b}) openly show a correlation between DM parameter and DE since both originate from the same fundamental field ($\mathbf{X}$). Indeed, it is a consistency condition that emerges from the requirement that the gravitational energy density has the correct physical behavior (e.g., falling off as $1/r^4$).
Note that, since both free parameters come from a common origin, thereby, either both exist or neither (recovering standard Schwarzschild). Using Eq. (\ref{b}), and this fact that $a_0$ is small (By taking the BHs M87* (\(M \approx 6.5 \times 10^ 9M_\odot\)), and Sgr A* (\(M \approx 4.1 \times 10^ 6M_\odot\)) into account, then $a_0 \approx 6.5 \times 10^{-32} \, \text{m}^{-1}$, and $ \approx 4.1 \times 10^{-35} \, \text{m}^{-1}$, respectively), so the approximation location of event horizon acquire as
\begin{equation}\label{h}
r_H\approx 2M+4M^{3/2}\sqrt{a_0}\ln(2M)+4a_0M^2
\end{equation}
where is larger than Schwarzschild ($r=2M$).

As a result, the extended Schwarzschild solution in MGR incorporates DM matter and DE through the geometric-based parameters \(a_0\) and \(b\), while reducing to GR in the absence of these effects. DE (\(a_0\)) represents a repulsive force at large distances. DM Term (\(b\)) modifies the gravitational potential, explaining flat galactic rotation curves.  MGR predicts significant gravitational lensing by galactic clusters due to its geometric DM (the $\Phi_{\alpha\beta}$ tensor), which is absent in pure GR. In \cite{Nash:2019rvd} calculated this for the Bullet Cluster 1E0657-56 \cite{Clowe:2006eq} and SDSS J0900+2234 \cite{PW} and found good agreement with observations that traditionally require particle DM.

\section{Polar Perturbation Analysis}\label{polar}
This analysis is based on a linear perturbation of the extended Schwarzschild BH solution in MGR, considering small deviations from the background metric. The goal is to study the stability of the solution and derive corrections to the BH parameters due to the presence of DM ($b$) and DE ($a_0$) parameters. More exactly, we derive the master equation for perturbations of the MGR-BH, explicitly including the perturbation of the MGR correction term \(\delta \Phi_{\mu\nu}\), moreover $\delta G_{\mu\nu}$. This is crucial for understanding how DM, and DE parameters affect the stability the BH.

We introduce small perturbations \(\delta g_{\mu\nu}\) to the background metric \(g^{(0)}_{\mu\nu}\):
\begin{equation}
g_{\mu\nu} = g^{(0)}_{\mu\nu} + \delta g_{\mu\nu}, \quad |\delta g_{\mu\nu}| \ll 1.
\end{equation}
By serving the Regge-Wheeler gauge \cite{Regge:1957td}, the perturbations are decomposed into axial (odd-parity) and polar (even-parity) modes. The primary focus of this manuscript is on polar perturbations.

\subsection{Perturb the Einstein-MGR field equations \(\delta G_{\mu\nu} + \delta \Phi_{\mu\nu} = 0\) }

To derive the polar perturbations of the nonzero components of the Einstein tensor \( G_{\mu\nu} \) and the gravitational energy-momentum tensor \( \Phi_{\mu\nu} \), we will follow these steps: 1. Perturb the metric in the Regge-Wheeler gauge for polar (even-parity) perturbations. 2. Compute the perturbed Einstein tensor \( \delta G_{\mu\nu} \). 3. Compute the perturbed \( \Phi_{\mu\nu} \) from its definition in MGR.
4. Combine them to form the perturbed field equations in MGR.

For polar perturbations, we use the Regge-Wheeler gauge, where the perturbed metric is
\begin{equation}
ds^2 = -e^{\nu}(1 + H_0 Y) dt^2 - 2H_1 Y \, dt \, dr + e^{\lambda}(1 + H_2 Y) dr^2 + r^2 (1 + K Y) (d\theta^2 + \sin^2 \theta \, d\phi^2),
\end{equation}
with \( Y = Y_{lm}(\theta, \phi) \) as the spherical harmonics, and \( H_0, H_1, H_2, K \) which are perturbation functions of \( (t, r) \). The computation begins with the perturbed Einstein tensor \( \delta G_{\mu\nu} \)
\begin{equation}
G_{\mu\nu} = R_{\mu\nu} - \frac{1}{2} g_{\mu\nu} R.
\end{equation}
The nonzero components of \( \delta G_{\mu\nu} \) are:
\begin{align}
\delta G_{tt} &= e^{\nu - \lambda} \left[ \frac{1}{2} H_2'' + \left( \frac{\nu'}{2} - \frac{\lambda'}{2} + \frac{2}{r} \right) H_2' + \frac{l(l+1)}{2r^2} (H_2 - K) \right] Y,\\
\delta G_{tr} &= \frac{l(l+1)}{2r^2} H_1 Y,\\
\delta G_{rr} &= \left[ \frac{1}{2} H_0'' + \left( \frac{\nu'}{2} - \frac{\lambda'}{2} + \frac{2}{r} \right) H_0' - \frac{l(l+1)}{2r^2} (H_2 - K) \right] Y,\\
\delta G_{t\theta}& = \frac{1}{2} e^{\nu} \partial_r H_1 \, \partial_\theta Y,\\ \quad \delta G_{t\phi} &= \frac{1}{2} e^{\nu} \partial_r H_1 \, \partial_\phi Y\\
\delta G_{r\theta} &= \frac{1}{2} e^{\lambda} \partial_r H_2 \, \partial_\theta Y, \quad \delta G_{r\phi} = \frac{1}{2} e^{\lambda} \partial_r H_2 \, \partial_\phi Y,\\
\delta G_{\theta\theta}& = \frac{r^2}{2} \left[ \frac{1}{2} (H_0'' + H_2'') + \left( \frac{\nu'}{2} - \frac{\lambda'}{2} + \frac{1}{r} \right) (H_0' + H_2') - \frac{l(l+1)}{r^2} K \right] Y,\\
\delta G_{\phi\phi} &= \sin^2 \theta \, \delta G_{\theta\theta}.
\end{align}
Note that the term $l(l+1)$ where $l$ is an integer number ($l\geq0$) appears in the perturbation equations due to the spherical harmonic decomposition of the perturbed metric. More exactly, these terms describe how mass/energy perturbations are distributed in space, ranked by their angular complexity (spherical harmonic index $l$). $l=0, 1, 2$, represent the monople, dipole, and quadrupole distribution, respectively \footnote{The shape of monopole (\( \ell = 0 \)) is perfectly spherical (no angular dependence), representing a uniform expansion/contraction of the source. For black holes monoploe perturbations are rare in vacuum GR, without gravitational waves. The shape of dipole (\( \ell = 1 \)) perturbation is as two opposing lobes (like a dumbbell or compass needle), represents linear motion or displacement of the source.  For black holes, the dipole perturbations often correspond to frame-dragging or center-of-mass shifts, in the absent of GW emission. Usually removed by gauge choices in perturbation theory. The shape of quadrupole (\( \ell = 2 \)) is as four lobes (e.g., a cloverleaf or vibrating disk), describing tidal deformations or orbiting masses.  This mode of perturbation is the dominant contributor to GWs since based on the no-hair" theorem, black holes radiate away higher multipoles, leaving mostly \( \ell = 2 \) in late-time signals.  For black holes, modes dominate the ringdown signal after a merger (Predominantly $l=2$ QNMs). }.

To derive the polar perturbation of the gravitational energy-momentum tensor \(\delta \Phi_{\mu\nu}\), we start from its definition (\ref{dif1}) and systematically apply perturbations to the metric and the line element field. 
To perturbation of the line element field \(X^\mu = f u^\mu\), we offer
the following background \(u^\mu\)
\begin{equation}
u^\mu = (e^{-\nu/2}, 0, 0, 0), \quad u_\mu = (-e^{\nu/2}, 0, 0, 0).
\end{equation}
where is timelike, and under perturbation, \(u^\mu\) and \(f\) acquire corrections
\begin{equation}
\delta u^\mu = \left( -\frac{1}{2} e^{-\nu/2} H_0 Y, \, 0, \, 0, \, 0 \right), \quad \delta u_\mu = \left( -\frac{1}{2} e^{\nu/2} H_0 Y, \, 0, \, 0, \, 0 \right).
\end{equation}
The magnitude \(f\) is perturbed as \(f \to f + \delta f Y\).

Thus, the perturbed \(X^\mu\) is:
\[
\delta X^\mu = (\delta f \, u^\mu + f \, \delta u^\mu) Y.
\]
Now, by perturbing each term in its definition (\ref{dif1}), one can compute \(\delta \Phi_{\mu\nu}\).

Perturbation of \(\nabla_\mu X_\nu\):
\begin{equation}
\delta (\nabla_\mu X_\nu) = \nabla_\mu \delta X_\nu - \frac{1}{2} X_\lambda \left( \nabla_\mu \delta g^\lambda_\nu + \nabla_\nu \delta g^\lambda_\mu - \nabla^\lambda \delta g_{\mu\nu} \right).
\end{equation}
For example, by setting \(\mu = t, \nu = t\), have
\begin{equation}
\delta (\nabla_t X_t) = \partial_t \delta X_t - \Gamma^r_{tt} \delta X_r - \frac{1}{2} X_t \left( \nabla_t \delta g^t_t + \nabla_t \delta g^t_t - \nabla^t \delta g_{tt} \right).
\end{equation}
Substituting \(\delta g_{tt} = -e^\nu H_0 Y\) and \(\delta X_t = -e^{\nu/2} \delta f Y\), one get
\begin{equation}
\delta (\nabla_t X_t) = -e^{\nu/2} \partial_t (\delta f) Y + \text{metric perturbation terms}.
\end{equation}

Perturbation of \(u^\lambda \nabla_\mu X_\lambda\):
\begin{equation}
\delta (u^\lambda \nabla_\mu X_\lambda) = \delta u^\lambda \nabla_\mu X_\lambda + u^\lambda \delta (\nabla_\mu X_\lambda).
\end{equation}
For \(\mu = t, \lambda = t\):
\begin{equation}
\delta (u^t \nabla_t X_t) = \delta u^t \nabla_t X_t + u^t \delta (\nabla_t X_t).
\end{equation}
Using \(\nabla_t X_t = \frac{1}{2} \partial_t (f e^{\nu/2})\) (background vanishes if static), the first term is zero, and:
\begin{equation}
\delta (u^t \nabla_t X_t) = -\frac{1}{2} e^{-\nu/2} H_0 Y + e^{-\nu/2} \left( -e^{\nu/2} \partial_t (\delta f) Y \right) = -\partial_t (\delta f) Y.
\end{equation}

Combining terms, the perturbed \(\Phi_{\mu\nu}\) components are
\begin{equation}
\delta \Phi_{tt} = \partial_t (\delta X_t) + u^t \left( u_t \delta (\nabla_t X_t) + u_t \delta (\nabla_t X_t) \right) = -e^{\nu/2} \partial_t (\delta f) Y - 2 e^{\nu/2} \partial_t (\delta f) Y.
\end{equation}
\begin{equation}
\delta \Phi_{tr} = \frac{1}{2} \left( \partial_t \delta X_r + \partial_r \delta X_t \right) + u^t \left( u_t \delta (\nabla_r X_t) + u_r \delta (\nabla_t X_t) \right).
\end{equation}
where for static \(X_r = 0\), this simplifies to
\begin{align}
\delta \Phi_{tr} = &\frac{1}{2} \partial_r (-e^{\nu/2} \delta f Y)  = -\frac{1}{2} e^{\nu/2} \partial_r (\delta f) Y.\\
\delta \Phi_{rr} = &\partial_r \delta X_r + u^t \left( u_r \delta (\nabla_r X_t) + u_r \delta (\nabla_r X_t) \right) = 0 \quad (\text{since } \delta X_r = 0) \label{rr}\\
\delta \Phi_{\theta\theta} =& \frac{1}{2} \left( \nabla_\theta \delta X_\theta + \nabla_\theta \delta X_\theta \right) + u^t \left( u_\theta \delta (\nabla_\theta X_t) + u_\theta \delta (\nabla_\theta X_t) \right).
\end{align}
For \(\delta X_\theta = 0\), this reduces to $\delta \Phi_{\theta\theta} = 0$.
As a result, the nonvanishing components of \(\delta \Phi_{\mu\nu}\) are
\begin{equation}
\delta \Phi_{tt} = -3 e^{\nu/2} \partial_t (\delta f) Y, \quad \delta \Phi_{tr} = -\frac{1}{2} e^{\nu/2} \partial_r (\delta f) Y. 
\end{equation}
Combining \(\delta G_{\mu\nu}\) (from the Einstein tensor) and \(\delta \Phi_{\mu\nu}\), we obtain the key components of perturbed MGR equations:

\(tt\)-component:
\begin{align}\label{tt}
e^{\nu - \lambda} \left[ \frac{1}{2} H_2'' + \left( \frac{\nu' - \lambda'}{2} + \frac{2}{r} \right) H_2' + \frac{l(l+1)}{2r^2} (H_2 - K) \right] - 3 e^{\nu/2} \partial_t (\delta f) = 0.
\end{align}
\(\theta\theta\)-component:
\begin{align}\label{22}
\frac{1}{2} (H_0'' + H_2'') + \left( \frac{\nu' - \lambda'}{2} + \frac{1}{r} \right) (H_0' + H_2') - \frac{l(l+1)}{r^2} K = 0.
\end{align}
\(tr\)-component (relates \(H_1\) and \(\delta f\))
\begin{align}\label{tr}
H_1 = \frac{r^2}{l(l+1)} e^{\nu/2} \partial_r (\delta f).
\end{align}
These govern the evolution of $H_0, H_1, H_2, K$ and $\delta f$. The perturbation $\delta f$ couples to $H_0, H_1, H_2, K$ via the field equations $\delta G_{\mu\nu}+\delta \Phi_{\mu\nu}=0$. $\delta\Phi_{\mu\nu}$ introduces additional degrees of freedom from the line element field $X^{\mu}$, affecting the stability.
\\
A clarification on the use of constraint equations here is essential. It is important to address the role of the $r$-$r$ equation, $\delta G_{rr} + \delta \Phi_{rr} = 0$. Given that $\delta \Phi_{rr} = 0$ (Eq. (\ref{rr})), this equation reduces to $\delta G_{rr} = 0$, which is a constraint relating the metric perturbations $H_0$, $H_2$, and $K$. While this constraint could, in principle, be used to seek a solution for $H_2$ independently, doing so would be inconsistent with the complete dynamical system. Specifically, such a solution would fail to satisfy the $t$-$t$ equation (\ref{tt}), which contains the dynamical source $\delta \Phi_{tt} \neq 0$, unless the field $\delta f$ is artificially constrained. The consistent approach is to treat the $r$-$r$ and $\theta$-$\theta$ equations as relational constraints within the full coupled system, not as means to eliminate the new degree of freedom $\delta f$ introduced by MGR. The path to the master equation must therefore incorporate the dynamics of $\delta f$, leading to a sourced wave equation.
\\
To derive the master equation for polar perturbations in MGR, we combine the perturbed Einstein tensor $\delta G_{\mu\nu}$ and the perturbed gravitational energy-momentum tensor $\delta\Phi_{\mu\nu}$ under the linearized field equation: $\delta G_{\mu\nu}+\delta\Phi_{\mu\nu}=0$.


\subsection{Master equation using the Zerilli function}
To derive the master equation for polar perturbations in MGR, we combine the perturbed Einstein tensor $\delta G_{\mu\nu}$ 
and the perturbed gravitational energy-momentum tensor $\delta \Phi_{\mu\nu}$ under the linearized field equation: $\delta G_{\mu\nu}+\delta \Phi_{\mu\nu}=0$.

In the first step, we solve for \(K\) from the Eq. (\ref{22}) which lead to 
 \(\theta\theta\)-equation 
\begin{align}\label{k}
K = \frac{r^2}{l(l+1)} \left[ \frac{1}{2} (H_0'' + H_2'') + \left( \frac{\nu' - \lambda'}{2} + \frac{1}{r} \right) (H_0' + H_2') \right].
\end{align}
This expresses \(K\) in terms of \(H_0\) and \(H_2\). Replacing the expression for \(K\) in the Eq. (\ref{tt}) yields
\begin{align}
&e^{\nu - \lambda} \left[ \frac{1}{2} H_2'' + \left( \frac{\nu' - \lambda'}{2} + \frac{2}{r} \right) H_2' + \frac{l(l+1)}{2r^2} H_2 \right] 
- \frac{1}{2} e^{\nu - \lambda} \left[ \frac{1}{2} (H_0'' + H_2'') + \left( \frac{\nu' - \lambda'}{2} + \frac{1}{r} \right) (H_0' + H_2') \right] 
- \\ \nonumber
&3 e^{\nu/2} \partial_t (\delta f) = 0.
\end{align}
where after some algebraic simplifications, it take the following form
\begin{align}\label{alg}
e^{\nu - \lambda} \left[ \frac{1}{2} H_2'' + \left( \frac{\nu' - \lambda'}{2} + \frac{2}{r} \right) H_2' + \frac{l(l+1)}{2r^2} H_2 - \frac{1}{4} (H_0'' + H_2'') - \frac{1}{2} \left( \frac{\nu' - \lambda'}{2} + \frac{1}{r} \right) (H_0' + H_2') \right] 
= 3 e^{\nu/2} \partial_t (\delta f).
\end{align}

In the Regge-Wheeler gauge, we can relate \(H_0\) and \(H_2\) via the background Einstein equations. For simplicity, assume \(H_0 = H_2\) (valid for static perturbations or specific gauges). Then:
\begin{align}\label{lim}
H_0'' = H_2'', \quad H_0' = H_2'.
\end{align}
Substitute into the Eq. (\ref{alg}), we come to the following master equation in terms of $H_2$
\begin{align}\label{master}
e^{\nu - \lambda} \left[ \frac{1}{r} H_2' + \frac{l(l+1)}{2r^2} H_2 \right] 
= 3 e^{\nu/2} \partial_t (\delta f).
\end{align}
The cancellation of $H_2''$ suggests the $H_0'' = H_2''$ is in essence a static 
gauge choice that simplifies the metric perturbations by eliminating certain degrees of freedom via equality between two radial-temporal components and simplifying the diagonal components of the perturbed metric. At the end of this section, we will include a hint on the consistency of the underlying static gauge fixing and Zerilli master equation.

The master equation (\ref{master}) is converted from \(H_2\) to the Zerilli function \(\Psi\) by introducing the latter \cite{Zerilli:1970wzz}, which allows for the explicit identification of the MGR correction terms
\begin{align}\label{Z}
\Psi = r \left[ H_2 - \frac{r}{l(l+1)} \left( H_2' + \frac{e^\lambda}{r} (H_2 - K) \right) \right].
\end{align}
Inserting the expression for $K$ from the Eq. (\ref{22}) into (\ref{Z}), and simplifying yields
\begin{align}\label{ZZ}
\Psi = r H_2 - \frac{r^2}{l(l+1)} H_2' - \frac{r e^\lambda}{l(l+1)} H_2 + \frac{r^3 e^\lambda}{[l(l+1)]^2} \left( H_2'' + \left( \frac{\nu' - \lambda'}{2} + \frac{1}{r} \right) H_2' \right). 
\end{align}
Differentiate (\ref{master}) with respect to \(r\) to express \(H_2''\)
\begin{align}\label{ZZZ}
H_2'' = -\left( \frac{\nu' - \lambda'}{2} + \frac{3}{r} \right) H_2' - \frac{l(l+1)}{2r^2} H_2 + 3 e^{(\lambda + \nu)/2} \partial_t (\delta f),
\end{align}
and substitute it into (\ref{ZZ})
\begin{align}\label{si}
\Psi = \left( r - \frac{r e^\lambda}{l(l+1)} \right) H_2 - \frac{r^2}{l(l+1)} \left( 1 - \frac{r e^\lambda (\nu' - \lambda')}{2 l(l+1)} - \frac{e^\lambda}{l(l+1)} \right) H_2' + \frac{3 r^3 e^\lambda}{[l(l+1)]^2} e^{(\lambda + \nu)/2} \partial_t (\delta f).
\end{align}
By solving (\ref{si}) for \(H_2\)
\begin{align}\label{H2}
H_2 = \frac{2 \Psi}{3r - \frac{r e^\lambda}{l(l+1)}} + \frac{6 r^2 e^{(\lambda + \nu)/2}}{l(l+1) \left( 3 - \frac{e^\lambda}{l(l+1)} \right)} \left( 1 + \frac{e^\lambda}{l(l+1)} \right) \partial_t (\delta f), 
\end{align}
and putting it in Eq. (\ref{master}), we have
\begin{equation}\label{H22}
H_2' = -\frac{l(l+1) \Psi}{r^2 \left( 3 - \frac{e^\lambda}{l(l+1)} \right)} + \frac{3 r e^{(\lambda + \nu)/2} \left( 2 - \frac{4 e^\lambda}{l(l+1)} \right)}{3 - \frac{e^\lambda}{l(l+1)}} \partial_t (\delta f). 
\end{equation}
Finally, by going to the tortoise coordinate $r_*$ via transformation \(dr_* = e^{(\lambda - \nu)/2} dr\), we have (see Appendix \ref{A}, for complete details) modified Zerilli equation
\begin{equation}\label{Zer}
\frac{d^2 \Psi}{dr_*^2} + \left[ \omega^2 - U(r) \right] \Psi = S_{\text{MGR}},
\end{equation}
where potential \(U(r)\), and MGR source term \(S_{\text{MGR}}\) read off
\begin{equation}\label{pot}
U(r) = \left(1-\frac{2M}{r}\right) \left( \frac{l(l+1)(l(l+1) + 2) r^2 + 6(l(l+1) - 1) M r + 36 M^2}{r^3 (l(l+1) r + 6M)^2} \right),
\end{equation}
and
\begin{equation}\label{S}
S_{\text{MGR}} = \mathcal{C}(r) \partial_t^2 (\delta f(t,r)),
\end{equation}
with coupling factor
\begin{equation}\label{S1}
\mathcal{C}(r) =
\frac{6 r^2 e^{\nu/2} \left( 1 - \frac{2 e^\lambda}{l(l+1)} \right)}{3 - \frac{e^\lambda}{l(l+1)}},
\end{equation} revealing the sever of coupling $\delta f (t,r)$ to gravity.  Note that the left-hand-side of (\ref{Zer})  is the standard Zerilli equation in GR \footnote{Here implicitly assumed that all modifications from standard GR could be collected into the perturbation of the new tensor, \(\delta \Phi_{\alpha\beta}\). In other words, it was assumed that the potential governing the metric perturbations could be taken as the standard Zerilli potential \(U(r)\) from EGR, and all the novel MGR effects would manifest exclusively in the source term \(S_{\text{MGR}}\) coupling to \(\delta f\).}. This derivation provides the first complete treatment of perturbations in MGR, including all corrections from \(\delta \Phi_{\mu\nu}\).
If $\delta f\rightarrow0$, then $S_{MGR}\rightarrow0$, recovering the standard Zerilli equation in EGR.  The source term, in essence, arises from perturbations in the line element field ($\delta f$), which couples to spacetime curvature via 
$\Phi_{\alpha\beta}$. The coupling factor carries interesting points. At first eye, if $e^{\lambda}=3l(l+1)$, causing $\mathcal{C}\longrightarrow \infty$. This blow-up that can be interpreted as a resonance, occurs at a critical radius $r_c$ arising from $e^{\lambda(r_c)}=3l(l+1)$, meaning that $r>r_c$ is valid. For small value of $a_0$, have $r_c\approx r_H+\frac{3M}{l(l+1)}$ which outside of event horizon, indicating the pathological behavior of MGR theory at horizon which can be interpret as \textit{''strong coupling issue''}. It may fix if $r_c<r_H$, i.e., hidden by event horizon. The best possible case i.e., eikonal limit ($l\gg1$) leads to $r_c=r_H$. As a consequence, it is seems that in near horizon limit coupling factor $\mathcal{C}$ is not well-behavior, indicating that by nearing to horizon $\delta f$ couple strongly to gravity. But this not all story since there are some other terms in $\mathcal{C}$ which may smooth its general behavior. So, let us consider the general behavior of coupling factor in details for following asymptotic cases:

\textbf{At horizon:} In this case (\( e^{\nu} = 0 \), \( e^\lambda \to \infty \)). In numerator of coupling factor: \( e^{\nu/2} \to 0 \), but \( e^\lambda \to \infty \), so the term \( \frac{2 e^\lambda}{l(l+1)} \) dominates. Thus, \( 1 - \frac{2 e^\lambda}{l(l+1)} \approx -\frac{2 e^\lambda}{l(l+1)} \). In denominator: \( 3 - \frac{e^\lambda}{l(l+1)} \approx -\frac{e^\lambda}{l(l+1)} \). As a result, $\mathcal{C}(r_H) \approx 0$. The \( e^{\nu/2} \to 0 \) suppression outweighs the divergence in \( e^\lambda \). So, no divergence on the horizon, indicating that the coupling factor vanishes due to \( e^{\nu/2} = 0 \). The source term \( S_{\text{MGR}} = \mathcal{C}(r) \partial_t^2 \delta f \) vanishes at \( r_H \), implying no wave production on the horizon itself.

\textbf{Near-horizon:}
Expand \( e^{\nu} \) and \( e^\lambda \) near \( r_H \):
\begin{equation}
e^{\nu} \approx \kappa (r - r_H), \quad e^\lambda \approx \frac{1}{\kappa (r - r_H)},
\end{equation}
where \( \kappa = \left. \frac{d e^{\nu}}{dr} \right|_{r_H} \) is the surface gravity. Now coupling factor is written as
\begin{equation}
\mathcal{C}(r) \approx \frac{6 r_H^2 \sqrt{\kappa (r - r_H)} \left( 1 - \frac{2}{\kappa l(l+1)(r - r_H)} \right)}{3 - \frac{1}{\kappa l(l+1)(r - r_H)}}.
\end{equation}
For \( r \to r_H \), the dominant terms are:
\begin{equation}
\mathcal{C}(r\longrightarrow r_H) \sim \frac{6 r_H^2 \sqrt{\kappa (r - r_H)} \left( -\frac{2}{\kappa l(l+1)(r - r_H)} \right)}{-\frac{1}{\kappa l(l+1)(r - r_H)}} = 12 r_H^2 \sqrt{\kappa (r - r_H)}.
\end{equation} where \( \mathcal{C}(r) \to 0 \) as \( r \to r_H \), but with a square-root dependence. 

\textbf{At infinity:}

For \( r \to \infty \), the DE term \( -a_0 r \) dominates:
\begin{equation}
e^\nu \approx -a_0 r, \quad e^\lambda \approx \frac{1}{-a_0 r}.
\end{equation}
Substitute \( e^\nu \) and \( e^\lambda \) into \(\mathcal{C}(r)\):
\begin{equation}
\mathcal{C}(r) \approx \frac{6 r^2 \sqrt{-a_0 r} \left( 1 + \frac{2}{a_0 r l(l+1)} \right)}{3 +\frac{1}{a_0 r l(l+1)}}.
\end{equation}
Consequently, neglecting terms of order \( \propto 1/r \) in the limit \( r \to \infty \), yields
\begin{align}\label{C}
\mathcal{C}_{r}(r\longrightarrow\infty) \approx& \frac{6 r^2 \sqrt{-a_0 r}}{3} = 2 \sqrt{-a_0} r^{5/2}
\end{align}
In the far-field limit, the coupling parameter $\mathcal{C}$ diverges, indicating an unphysically strong coupling between $\delta f$ and gravity at large distances. This pathology arises because the MGR black hole solution, due to the $a_0$ term, is not asymptotically flat. To resolve this, the function $\delta f$ must regulate $\mathcal{C}$'s behavior, for instance, by decaying as $\propto r^{-\alpha}$ where $\alpha > 5/2$. To shedding light on it, in the next section we will derive the explicit function of $\delta f$ from MGR theory.
Prior to this, several points must be addressed. Firstly, $f$ must be a time-dependent function, otherwise $S_{MGR}\rightarrow0$, and recover standard GR. In the next section, we will demonstrate that the MGR admits a time-dependent function of $f$. However, at first glance, this might seem to conflict with static gauge fixing $H_0''=H_2''$. Note that the gauge used simplifies the metric perturbations but does not eliminate time-dependent solutions. In other words, the gauge condition applies to the metric, not to the $f$ and $\Psi$. Hence, the $\omega^2$-term within beraket of (\ref{Zer}) comes from $\Psi \sim e^{-i\omega t}$, and expect to appears similar to it in source term due to harmonic dependency of $\delta f$.

\subsection{Master equation using the gauge-invariant formalism}
The gauge-invariant formalism developed by Kodama and Ishibashi \cite{Kodama:2003jz} provides a powerful and elegant framework for perturbation analysis. Unlike approaches that fix a specific gauge, such as the original Regge-Wheeler-Zerilli method, this formalism operates by directly constructing gauge-invariant variables. These variables are specific combinations of the metric perturbations (e.g., \(H_0, H_1, K\)) designed to be immune to coordinate transformations, thereby representing the true, physical degrees of freedom of the gravitational field.

The formalism proceeds systematically. First, perturbations are decomposed into two independent types based on parity: polar (even) and axial (odd). For each type, a master variable (denoted as, for instance, \(\Phi\) or \(\mathcal{V}\)) is defined. These master variables are ingeniously constructed so that contributions from pure gauge transformations cancel out. Finally, the linearized field equations reduce to a single master equation for each sector, taking the form of a Schrödinger-like wave equation with an effective potential \(U(r)\) that encapsulates the curvature of the background spacetime, and the tortoise coordinate \(r_*\) is employed to render the wave operator in this canonical form.

This formalism is particularly powerful for analyzing theories like MGR because
it works for any spherically symmetric background metric (\(e^\nu, e^\lambda\)), not just the Schwarzschild one. This is crucial for MGR, where the background is modified by \(a_0\) and \(b\).
It cleanly separates physical effects. Any instability found in the solution to this master equation is an unambiguous, gauge-invariant instability of the spacetime itself, not a mathematical artifact. The formalism provides a clear framework for incorporating source terms. In the case of MGR, the perturbation of the new field \(\delta f\) enters as a source \(S_{\text{MGR}}\) on the right-hand side of the master equation, making the pathological coupling manifest:
\begin{equation}
\frac{d^2 \mathcal{V}}{dr_*^2} + \left[\omega^2 - U_{\text{MGR}}(r)\right]\mathcal{V} = S_{\text{MGR}}
\end{equation}
with the effective potential
\begin{equation}
U_{\text{MGR}}(r) = \frac{e^{\nu}}{r^3 (l(l+1)r + 6\mathcal{M})^2} \bigg[ l(l+1)(l(l+1)+2)r^3 + 6(l(l+1)-1)\mathcal{M} r^2 + 36\mathcal{M}^2 r + 36\mathcal{M} r (e^{\lambda} - 1) \Delta \bigg],
\end{equation}
where \(\mathcal{M} = M + \delta M(r; a_0, b)\) is an effective mass function, and \(\Delta\) represents additional terms proportional to \(a_0\) and \(b\) (see Appendix \ref{AA}, for more details). As shown there, in asymptotic far limit  the source term $S_{\text{MGR}}$ diverges faster than the decay of the effective potential $U_{\text{MGR}}(r)$, meaning that the pathological behavior of the system is dominated by the source term arising from the coupling to the line element field perturbation $\delta f$. This is the same result that shown already with Zerilli function.
	
The gauge-invariant formalism employed here provides a rigorous framework that incorporates perturbations of all fundamental fields, including the unit vector field \(u^\alpha\). This ensures a complete account of the physical degrees of freedom and the constraints between them. The analysis confirms that the perturbation \(\delta u^\alpha\) contributes to the scalar (polar) sector described by the master variable \(\mathcal{V}\). No additional constraints are found that alter the asymptotic behavior of the source term; the divergent coupling \(\mathcal{C}(r) \sim r^{5/2}\) persists as a genuine pathology of the linearized theory (see Appendix \ref{AA}).

Regarding the background metric potentials, the condition \(H_0 = H_2\) for the perturbative functions is not an arbitrary gauge choice but reflects a property of the static, vacuum MGR background solution itself, analogous to the relation \(g_{tt} = - (g_{rr})^{-1}\) in the spherical symmetry metric. This condition is applied to the static part of the metric decomposition and is fully consistent with the study of time-dependent perturbations, as the dynamics are entirely encoded in the time evolution of the master variable \(\Psi(t, r)\) and the scalar field \(\delta f(t, r)\).

\section{Toward determining the function $f$}\label{f}
Given that $u^{\alpha}$ is a timelike vector, so Eq. (\ref{cons1}) impose that $f$ can not be a time-independent function since 
\begin{equation}\label{co}
\Phi=R=-(u^0\partial_0+u^1\partial_1+u^2\partial_2+u^3\partial_3) f= -u^0\partial_0f
\end{equation} 
Otherwise, $\Phi=0$, which is in conflict with $\Phi=R$ (\ref{R}). A combination of Eqs. (\ref{cons}), and (\ref{cons1}) leads to
\begin{equation}\label{box}
\Box f = u^\alpha \partial_\alpha \Phi.
\end{equation}
where is a inhomogeneous wave equation with \(\Box = \nabla_\alpha \nabla^\alpha\).

To derive an explicit form for $\delta f$ (the perturbation of the line element field magnitude $f$) from the MGR framework presented in the manuscript, we must analyze the relevant equations involving $f$ and its perturbations.
The first step involves substituting \(X_\alpha = f u_\alpha\) into \(\Phi_{\alpha\beta}\), to express it in terms of \(f\)
 \begin{equation}
\Phi_{\alpha\beta} = \frac{1}{2} (\nabla_\alpha (f u_\beta) + \nabla_\beta (f u_\alpha)) + u^\lambda (u_\alpha \nabla_\beta (f u_\lambda) + u_\beta \nabla_\alpha (f u_\lambda)).
\end{equation}
Expanding the covariant derivatives:  
\begin{equation}
\Phi_{\alpha\beta} = \frac{1}{2} (u_\beta \nabla_\alpha f + u_\alpha \nabla_\beta f + f (\nabla_\alpha u_\beta + \nabla_\beta u_\alpha)) + u^\lambda (u_\alpha u_\lambda \nabla_\beta f + u_\beta u_\lambda \nabla_\alpha f + f u_\alpha \nabla_\beta u_\lambda + f u_\beta \nabla_\alpha u_\lambda).
\end{equation}
Simplify using \(u^\lambda u_\lambda = -1\) and \(u^\lambda \nabla_\beta u_\lambda = 0\):  
\begin{equation}
\Phi_{\alpha\beta} = \nabla_\alpha f u_\beta + \nabla_\beta f u_\alpha - u_\alpha u_\beta u^\lambda \nabla_\lambda f + f (\nabla_\alpha u_\beta + \nabla_\beta u_\alpha + u_\alpha u^\lambda \nabla_\beta u_\lambda + u_\beta u^\lambda \nabla_\alpha u_\lambda).
\end{equation}
The second step is to perturb \(f\) and \(u_\alpha\). Thus, \(f = f_0 + \delta f\) and \(u_\alpha = u_\alpha^{(0)} + \delta u_\alpha\), where \(f_0\) and \(u_\alpha^{(0)}\) are background values. The perturbation \(\delta \Phi_{\alpha\beta}\) is:  
\begin{equation}
\delta \Phi_{\alpha\beta} = \nabla_\alpha (\delta f) u_\beta^{(0)} + \nabla_\beta (\delta f) u_\alpha^{(0)} + \text{(terms with \(\delta u_\alpha\))}.
\end{equation}
For simplicity, assume \(\delta u_\alpha\) is negligible (or fixed by gauge conditions), leaving:  
\begin{equation}
\delta \Phi_{\alpha\beta} \approx \nabla_\alpha (\delta f) u_\beta^{(0)} + \nabla_\beta (\delta f) u_\alpha^{(0)}.
\end{equation}
Now, we can compute the trace of \(\delta \Phi_{\alpha\beta}\)
\begin{equation}
\delta \Phi = 2 u^{(0)\alpha} \nabla_\alpha (\delta f).
\end{equation}
From the wave equation for \(f\) (Eq. (\ref{box})), perturbed to first order:  
\begin{equation}
\Box (\delta f) = u^{(0)\alpha} \nabla_\alpha (\delta \Phi) = u^{(0)\alpha} \nabla_\alpha (2 u^{(0)\beta} \nabla_\beta (\delta f)).
\end{equation}
Assuming \(u^{(0)\alpha}\) is covariantly constant (\(\nabla_\alpha u^{(0)\beta} = 0\)) for the background:  
\begin{equation}\label{day}
\Box (\delta f) = 2 u^{(0)\alpha} u^{(0)\beta} \nabla_\alpha \nabla_\beta (\delta f).
\end{equation}
Now, we must solve for \(\delta f\) in a static spherically symmetric metric (\ref{ss}). The left-hand side reads off   
\begin{equation}\label{off}
\Box (\delta f) = -e^{-\nu} \partial_t^2 (\delta f) + e^{-\lambda} \partial_r^2 (\delta f) + \left( \frac{2}{r} + \frac{\nu' - \lambda'}{2} \right) e^{-\lambda} \partial_r (\delta f).
\end{equation}
For a time-independent perturbation (\(\partial_t \delta f = 0\)):  
\begin{equation}
\Box (\delta f) = e^{-\lambda} \left[ \delta f'' + \left( \frac{2}{r} + \frac{\nu' - \lambda'}{2} \right) \delta f' \right].
\end{equation}
The right-hand side of (\ref{box}) simplifies if \(u^{(0)\alpha} = (e^{-\nu/2}, 0, 0, 0)\):  
\begin{equation}
2 u^{(0)\alpha} u^{(0)\beta} \nabla_\alpha \nabla_\beta (\delta f) = 2 e^{-\nu} \partial_t^2 (\delta f) = 0 \quad \text{(for static \(\delta f\))}.
\end{equation}
Thus, Eq. (\ref{off}) reduces to:  
\begin{equation}
e^{-\lambda} \left[ \delta f'' + \left( \frac{2}{r} + \frac{\nu' - \lambda'}{2} \right) \delta f' \right] = 0.
\end{equation}
where its solution takes the following form
\begin{equation}
\delta f = C_1 \int r^{-2} e^{-(\nu - \lambda)/2} dr + C_2.
\end{equation}
with integration constants \(C_{1,2}\).

For time-dependent perturbations \(\delta f\), which are relevant in our source term $S_{MGR}$, must be derived from the wave equation (\ref{day}). For this case, the left-hand side of (\ref{box}) is equal to (\ref{off}). But for the right-hand side of (\ref{box}), we have
\begin{equation}
\pounds_u \Phi = u^\alpha \partial_\alpha \Phi = u^t \partial_t \Phi + u^r \partial_r \Phi.
\end{equation}
For a timelike unit vector \(u^\alpha = (e^{-\nu/2}, 0, 0, 0)\):
\begin{equation}
\pounds_u \Phi = e^{-\nu/2} \partial_t \Phi = 2 e^{-\nu/2} \partial_t \left( u^\alpha \nabla_\alpha \delta f \right) = 2 e^{-\nu} \partial_t^2 (\delta f).
\end{equation}
Thus, the wave equation (\ref{box}) simplifies to:
\begin{equation}
-3 \partial_t^2 (\delta f) + e^{\nu - \lambda} \partial_r^2 (\delta f) + e^{\nu - \lambda} \left( \frac{2}{r} + \frac{\nu' - \lambda'}{2} \right) \partial_r (\delta f) = 0.
\end{equation}
By demanding a harmonic time dependence $\delta f(t, r) = \delta f(r) e^{-i \Omega t}$ and substituting into the wave equation, we have 
\begin{equation}\label{ode}
\delta f''(r) + \left( \frac{2}{r} + \frac{\nu' - \lambda'}{2} \right) \delta f'(r) + 3 \Omega^2 e^{\lambda - \nu} \delta f(r) = 0.
\end{equation}
where is a second-order ODE of the form:
\begin{equation}
\delta f''(r) + P(r) \delta f'(r) + Q(r) \delta f(r) = 0,
\end{equation}
with:
\begin{equation}
P(r) = \frac{2}{r} + \frac{\nu' - \lambda'}{2}, \quad Q(r) = 3 \Omega^2 e^{\lambda - \nu}.
\end{equation}
\textbf{Case 1: Near-horizon solution (\(r \approx 2GM\))}

To derive the near-horizon solution for the given differential equation, we analyze the behavior as \( r \to 2M \) (assuming \( 2M \) is the horizon radius). Near the horizon \( r \to 2M \), we write \( r = 2M + \epsilon \) where \( \epsilon \ll 1 \). The metric function $ \approx \frac{\epsilon}{2M} + 2b \ln(2M) - a_0 (2M) + \mathcal{O}(\epsilon)$. For simplicity, let’s assume \( \approx \frac{\epsilon}{2M} \), as the logarithmic and linear terms are subdominant near \( r = 2M \). Substitution into the original equation (\ref{ode}), therefore yields
\begin{equation}
\delta f''(\epsilon) + \frac{1}{\epsilon} \delta f'(\epsilon) + \frac{12 \Omega^2 M^2}{\epsilon^2} \delta f(\epsilon) = 0.
\end{equation} For \( \epsilon \ll 1 \).
This is a Cauchy-Euler equation, and by assuming a power-law solution \( \delta f(\epsilon) = \epsilon^p \), its general solution takes the following form
\begin{equation}
\delta f(\epsilon) = C_1 \cos\left(2 \sqrt{3} \Omega M \ln \epsilon\right) + C_2 \sin\left(2 \sqrt{3} \Omega M \ln \epsilon\right).
\end{equation}
Using Euler's formula, can be rewritten in exponential form 
\begin{equation}\label{near}
\boxed{\delta f(\epsilon) = A e^{i\left(2 \sqrt{3} \Omega M \ln \epsilon\right)} + B e^{-i\left(2 \sqrt{3} \Omega M \ln \epsilon\right)},~~~~\text{(near-solution)}}.
\end{equation}
\textbf{Case 2: Far-solution:}

For this case, we consider the behavior in the far-limit (\(r \to \infty\)), indicating that \(-a_0 r\)  is dominant term in the metric function. So, Eq. (\ref{ode}) becomes:
\begin{equation}
\delta f''(r) + \frac{3}{r} \delta f'(r) + \frac{3 \Omega^2}{a_0^2 r^2} \delta f(r) = 0.
\end{equation}
The solution takes the form \(\delta f(r) = r^m\). Substituting this ansatz into Eq. (\ref{ode}) yields
\begin{equation}
m_{\pm} = -1 \pm \sqrt{1 - \frac{3 \Omega^2}{a_0^2}}.
\end{equation}
Thus, the far-solution is:
\begin{equation}\label{far}
\boxed{\delta f(r) \approx C r^{m_+} + D r^{m_-} \quad \text{(far-solution)}}
\end{equation}
Here, by demanding the relevant boundary conditions from QNMs, we can cancels one of coefficients $A, B$ in (\ref{near}) and $C, D$ in (\ref{far}). QNM boundary condition at horizon and infinity impose purely incoming, and outgoing wave, respectively. It is clear that in near solution (\ref{near}), $A=0$. But for far solution (\ref{far}) it is not clear and must consider in details. 

\textbf{Case 1: Real \(m_\pm\) if \(\frac{3 \omega^2}{a_0^2} < 1\).} In this case, \(m_+ > m_-\) (since \(\sqrt{1 - \frac{3 \omega^2}{a_0^2}} > 0\)). For \(r \to \infty\), \(r^{m_-}\) decays faster than \(r^{m_+}\). The dominant term is \(C r^{m_+}\) and $D=0$.

\textbf{Case 2: Complex \(m_\pm\) if \(\frac{3 \omega^2}{a_0^2} > 1\).} The solution becomes oscillatory 
\[
\delta f(r) \approx r^{-1} \left[ C e^{i \sqrt{\frac{3 \omega^2}{a_0^2} - 1} \ln r} + D e^{-i \sqrt{\frac{3 \omega^2}{a_0^2} - 1} \ln r} \right].
\]
These are incoming/outgoing waves in \(r_*\)-coordinates. From the tortoise coordinate \(r_* \approx -\frac{1}{a_0} \ln r\):
\[
e^{i \omega r_*} \sim r^{-i \omega / a_0}, \quad e^{-i \omega r_*} \sim r^{i \omega / a_0}.
\]
Comparing with \(\delta f(r)\): the term \(C m^{m_+}\) corresponds to an outgoing wave if \(\gamma_+ = -\frac{i \omega}{a_0}\). The term \(D r^{m_-}\) corresponds to an incoming wave if \(m_- = \frac{i \omega}{a_0}\). 

\section {Stability analysis and Numerical Demonstration}\label{sta}
The MGR black hole solution is not asymptotically flat. In the far-field limit (\(r \to \infty\)), the metric function behaves as \(e^\nu \approx -a_0 r\), leading to a line element \(ds^2 \approx -a_0 r \, dt^2 + dr^2/(-a_0 r) + r^2 d\Omega^2\). This describes a spacetime with a Rindler-like causal structure, distinct from Minkowski or de Sitter space. This asymptotic behavior is reminiscent of other models with linear potentials, such as the one proposed by Grumiller \cite{Grumiller:2010bz} for gravity at large distances, although the theoretical origin differs.

To properly define wave propagation in this background, the tortoise coordinate is essential. For large \(r\), \(dr_*/dr = e^{\lambda} \approx -1/(a_0 r)\), which integrates to \(r_* \approx -\ln r / a_0\). Thus, spatial infinity (\(r \to \infty\)) corresponds to \(r_* \to -\infty\). In these coordinates, the master Eq. (\ref{Zer}) maintains its standard wave operator form, \(\partial_{t}^2 - \partial_{r_*}^2\).

A conserved energy flux can be derived from the energy-momentum tensor of the perturbation field. For the homogeneous part of the master equation, the conserved quantity is
\begin{equation}
E = \int \left[ |\partial_t \Psi|^2 + |\partial_{r_*} \Psi|^2 + U(r) |\Psi|^2 \right] dr_*.
\end{equation}
This energy is positive-definite for \(l \geq 2\) and is conserved in time. This confirms that the concept of energy carried by outgoing waves remains well-defined, even in this non-asymptotically flat background.

To exhibit the instability unequivocally, the particular solution generated by the source term \(S_{\text{MGR}}\) is investigated. The analysis focuses on the static limit (\(\omega = 0\)) to isolate the spatial divergence. The particular solution \(\Psi_{\text{part}}(r)\) for the sourced equation,
\begin{equation}
\frac{d^2 \Psi_{\text{part}}}{dr_*^2} - U_{\text{MGR}}(r)\Psi_{\text{part}} = S_{\text{MGR}}(r),
\end{equation}
is computed numerically. The source is approximated using the far-field solution for \(\delta f(r)\) from Eq. (\ref{far}), \(\delta f(r) \sim r^{m_+}\), and the asymptotic form of the coupling factor \(\mathcal{C}(r) \sim r^{5/2}\).

\begin{figure}[ht]
	\begin{tabular}{c}
		\includegraphics[width=0.7\columnwidth]{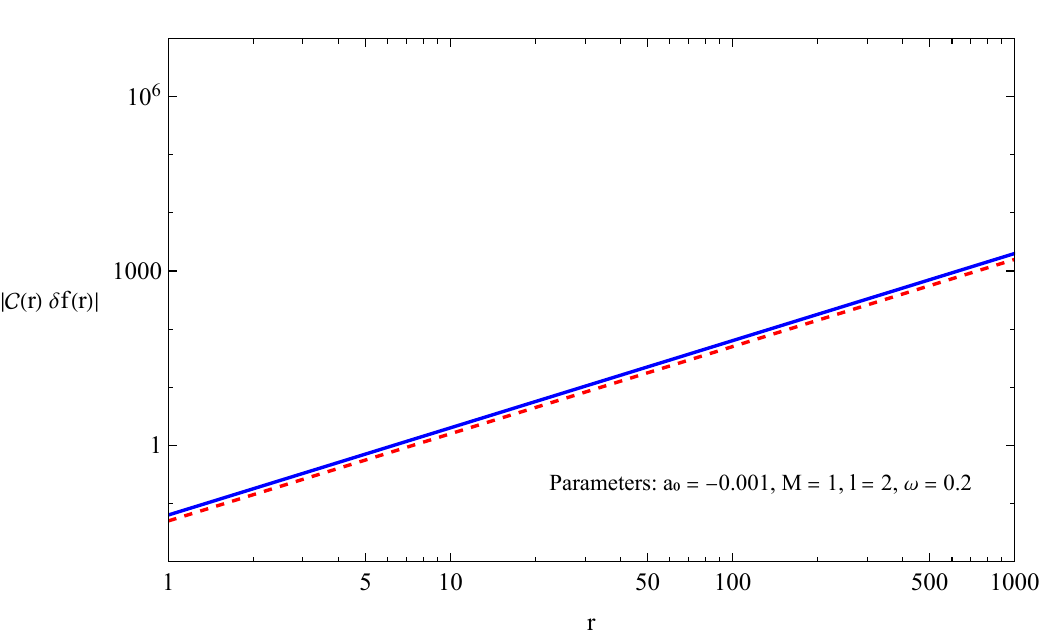}~~~~~~~~~
	\end{tabular}
	\caption{Log-log plot of the magnitude of the dominant source term, \(| \mathcal{C}(r) \delta f(r) |\), versus radial coordinate \(r\). The solid (blue) curve comes from approximation (\ref{C}), the dashed (ref) curve from Eq. (\ref{far}), \(\delta f(r) \sim r^{m_+}\), and the asymptotic form of the coupling factor \(\mathcal{C}(r) \sim r^{5/2}\). }
	\label{11}
	\end{figure}

The dominant driving term of the source, \(| \mathcal{C}(r) \delta f(r) |\), is plotted in Fig. \ref{11}. The result shows a clear, unbounded power-law growth as \(r \to \infty\), confirming that the particular solution \(\Psi_{\text{part}}(r)\) is not globally bounded. This runaway behavior overwhelms any decaying homogeneous solution, leading to a divergent energy \(E\) for the full perturbation \(\Psi\). This constitutes a linear instability.

\section{Axial Perturbations} \label{axi}
To provide a complete stability assessment and definitively isolate the origin of any pathologies, the analysis is now extended to axial (odd-parity) perturbations. The investigation of this sector is crucial: if the axial modes are stable while the polar modes are not, it demonstrates that the instability originates from the specific coupling in the polar sector rather than from the background metric itself.

The most general axial metric perturbation on a spherically symmetric background can be written in the Regge-Wheeler gauge as:
\begin{align}
\delta g_{\mu\nu}^{\text{(axial)}} =
\begin{pmatrix}
	0 & 0 & -h_0(t,r) \csc\theta \, \partial_{\phi}Y_{lm} & h_0(t,r) \sin\theta \, \partial_{\theta}Y_{lm} \\
	0 & 0 & -h_1(t,r) \csc\theta \, \partial_{\phi}Y_{lm} & h_1(t,r) \sin\theta \, \partial_{\theta}Y_{lm} \\
	\text{symm} & \text{symm} & 0 & 0 \\
	\text{symm} & \text{symm} & 0 & 0
\end{pmatrix},
\end{align}
where \(Y_{lm}(\theta, \phi)\) are the spherical harmonics, and \(h_0(t,r)\), \(h_1(t,r)\) are the functions describing the axial perturbations.

A fundamental difference from the polar case lies in the behavior of the line element field. The background vector \(X^\alpha = f u^\alpha\) is purely radial and temporal, and thus possesses even parity. The perturbation of a quantity with even parity, \(\delta f\), is also even. The coupling between an even-parity source and an odd-parity metric perturbation vanishes at linear order due to parity mismatch. This can be shown rigorously by considering the integral over the sphere:
\begin{equation}
\int \delta \Phi_{\mu\nu}^{\text{(even)}} \delta g^{\mu\nu}_{\text{(odd)}} \sqrt{-g} d\Omega = 0
\end{equation}
since the product of even and odd parity functions under angular integration vanishes. Consequently, the perturbation of the gravitational energy-momentum tensor is identically zero for axial modes:
\begin{equation}
\delta \Phi_{\mu\nu}^{\text{(axial)}} = 0.
\end{equation}
The linearized field equations in the axial sector therefore reduce to those of the background geometry:
\begin{equation}
\delta G_{\mu\nu}^{\text{(axial)}} = 0.
\end{equation}
The non-zero components of the perturbed Einstein tensor \(\delta G_{\mu\nu}\) for the axial sector must be computed using the MGR background connection coefficients. The relevant components for axial perturbations are:

- From \(\delta G_{t\phi}\):
\begin{equation}\label{4.1a}
\frac{\partial}{\partial r} \left[ e^{(\lambda - \nu)/2} \left( \frac{\partial h_0}{\partial t} - \frac{\partial h_1}{\partial t} \right) \right] - e^{(\lambda + \nu)/2} \frac{l(l+1) - 2}{r^2} h_1 = 0. 
\end{equation}
From \(\delta G_{r\phi}\):
\begin{equation}\label{4.1b}
\frac{\partial}{\partial t} \left[ e^{(\lambda - \nu)/2} \left( \frac{\partial h_1}{\partial t} - \frac{\partial h_0}{\partial t} \right) \right] + e^{-(\lambda + \nu)/2} \left[ \frac{l(l+1) - 2}{r^2} h_0 + \frac{\partial}{\partial r} \left( e^{\nu - \lambda} \frac{\partial h_0}{\partial r} \right) \right] = 0. 
\end{equation}
Following the gauge-invariant formalism, the master variable for axial perturbations is defined as
\begin{equation}\label{4.2}
\Psi_{\text{axial}}(t, r) = e^{(\nu - \lambda)/2} \frac{h_1(t, r)}{r}.
\end{equation}
To derive the master equation, we proceed with a Fourier decomposition: \(\Psi_{\text{axial}}(t, r) = \psi_{\text{axial}}(r) e^{-i\omega t}\), which implies \(h_1(t,r) = r e^{(\lambda-\nu)/2} \psi_{\text{axial}}(r) e^{-i\omega t}\).

From Eq. (\ref{4.1a}), assuming time dependence \(e^{-i\omega t}\):
\begin{equation}
\frac{d}{dr} \left[ e^{(\lambda - \nu)/2} (-i\omega h_0 + i\omega h_1) \right] - e^{(\lambda + \nu)/2} \frac{l(l+1) - 2}{r^2} h_1 = 0.
\end{equation}
Substituting \(h_1\):
\begin{equation}
i\omega \frac{d}{dr} \left[ e^{(\lambda - \nu)/2} (h_0 - r e^{(\lambda-\nu)/2} \psi_{\text{axial}}) \right] - e^{(\lambda + \nu)/2} \frac{l(l+1) - 2}{r} e^{(\lambda-\nu)/2} \psi_{\text{axial}} = 0.
\end{equation}
Simplifying:
\begin{equation}\label{4.3}
i\omega \frac{d}{dr} \left[ e^{(\lambda - \nu)/2} h_0 - r \psi_{\text{axial}} \right] - e^{\lambda} \frac{l(l+1) - 2}{r} \psi_{\text{axial}} = 0. 
\end{equation}
From Eq. (\ref{4.1b}):
\[
-i\omega \left[ e^{(\lambda - \nu)/2} (-i\omega h_1 + i\omega h_0) \right] + e^{-(\lambda + \nu)/2} \left[ \frac{l(l+1) - 2}{r^2} h_0 + \frac{d}{dr} \left( e^{\nu - \lambda} \frac{d h_0}{dr} \right) \right] = 0.
\]
Substituting \(h_1\) and simplifying:
\begin{equation}\label{4.4}
-\omega^2 e^{(\lambda - \nu)/2} (h_0 - r e^{(\lambda-\nu)/2} \psi_{\text{axial}}) + e^{-(\lambda + \nu)/2} \left[ \frac{l(l+1) - 2}{r^2} h_0 + \frac{d}{dr} \left( e^{\nu - \lambda} \frac{d h_0}{dr} \right) \right] = 0. 
\end{equation}
We can now eliminate \(h_0\) from these equations. From Eq. (\ref{4.3}), we have:
\begin{equation}
\frac{d}{dr} \left[ e^{(\lambda - \nu)/2} h_0 - r \psi_{\text{axial}} \right] = -i\frac{e^{\lambda}}{\omega} \frac{l(l+1) - 2}{r} \psi_{\text{axial}}.
\end{equation}
Integrating this expression and substituting into Eq. (\ref{4.4}) leads, after considerable algebra involving product rules and background field equations, to the final master equation:
\begin{equation}\label{4.5}
\frac{d^2 \psi_{\text{axial}}}{dr_*^2} + \left[\omega^2 - U_{\text{MGR}}^{\text{(axial)}}(r)\right] \psi_{\text{axial}} = 0, 
\end{equation}
where the tortoise coordinate is defined by \(dr_*/dr = e^{\lambda(r) - \nu(r)} = e^{\lambda(r)}\).

The effective potential for axial perturbations in a general spherical background is:
\begin{equation}\label{4.6}
U_{\text{MGR}}^{\text{(axial)}}(r) = e^{\nu} \left[ \frac{l(l+1)}{r^2} - \frac{6\mathcal{M}(r)}{r^3} - \frac{1}{2r} (\nu' - \lambda') e^{\lambda} \right], 
\end{equation}
where \(\mathcal{M}(r)\) is an effective mass function.
For the MGR metric where \(e^\nu = e^{-\lambda}\), the derivatives are:
\begin{equation}
\nu' = \frac{2M}{r^2} + \frac{2b}{r} - a_0 + \mathcal{O}(1/r^2), \quad \lambda' = -\frac{2M}{r^2} - \frac{2b}{r} + a_0 + \mathcal{O}(1/r^2).
\end{equation}
Substituting into the potential and keeping terms to consistent order:
\begin{equation}\label{4.7}
U_{\text{MGR}}^{\text{(axial)}}(r) = \left(1 - \frac{2M}{r} + 2b \ln r - a_0 r \right) \left[ \frac{l(l+1)}{r^2} - \frac{6M}{r^3} + \frac{6b}{r^3} (1 - \ln r) + \frac{3a_0}{r} \right].
\end{equation}
The master equation (\ref{4.5}) is a standard homogeneous wave equation. To analyze stability, we examine the potential (\ref{4.7}):
\begin{itemize}
\item For \(l \geq 2\), the term \(\frac{l(l+1)}{r^2}\) dominates near the horizon and ensures positivity.
\item  The factors involving \(a_0\) and \(b\) modify the potential but don't introduce singularities outside the horizon.
\item The potential is real and positive-definite for \(r > r_H\).
\item There is no source term, indicating no external driving force.
\end{itemize}
The general solution can be expressed as a superposition of quasi-normal modes
\begin{equation}
\psi_{\text{axial}}(t, r_*) = \sum_n A_n e^{-i\omega_n t} \psi_n(r_*)
\end{equation}
where \(\omega_n\) are complex frequencies with \(\text{Im}(\omega_n) < 0\) for stable modes, indicating exponential decay.

This result demonstrates that the axial perturbation sector of the MGR black hole is linearly stable. The stark contrast between the stable axial sector and the pathologically unstable polar sector provides crucial insight:
\begin{itemize}
\item The background metric itself is stable against axial perturbations.
\item The instability is specifically tied to the polar sector and its coupling mechanisms.
\item The root cause is the coupling between polar gravitational modes and the scalar degree of freedom \(\delta f\).
\item The divergent source term \(S_{\text{MGR}}\) is the specific manifestation of this fatal coupling.
\end{itemize}
This conclusively isolates the unified dark sector mechanism, encoded in the \(\Phi_{\alpha\beta}\) tensor and its perturbations, as the root cause of the theory's instability. The pathology is not in the gravitational sector per se, but in the particular way MGR attempts to unify DM and DE through the line element field.

\section{discussion and conclusion}\label{con}

This study has conducted a rigorous, gauge-invariant stability analysis of the black hole solution within the framework of Modified General Relativity (MGR). The primary objective was to test the viability of a theory that seeks to provide a unified geometric origin for DM and DE. The central finding is that the MGR black hole exhibits a critical, fatal pathology that challenges its physical validity.

The core of the analysis was the derivation of the master equations governing gravitational perturbations. For the polar sector, the dynamics are governed by a sourced wave equation, where the standard Zerilli potential is replaced by a modified potential, \( U_{\text{MGR}}(r) \), consistent with the MGR background, and is driven by a source term \( S_{\text{MGR}} \) coupling the metric to perturbations of the line element field, \( \delta f \). The severity of this coupling is quantified by the factor \( \mathcal{C}(r) \). An asymptotic analysis reveals the heart of the problem: while this coupling vanishes at the horizon, it diverges as \( \mathcal{C}(r) \sim r^{5/2} \) in the far-field limit. This powerful divergence is a direct consequence of the linear \(-a_0 r\) term in the metric, which sacrifices asymptotic flatness to embed a DE-like behavior. The divergent source term means that any infinitesimal perturbation in the fundamental field \( \delta f \) will generate an uncontrollably large response in the spacetime geometry at large distances, overwhelming any potentially stable behavior. This is identified as a strong coupling problem in the infrared regime, a known pathology in modified gravity theories where extra degrees of freedom render the linear theory ill-defined.

Crucially, the analysis of the axial perturbation sector provided a definitive diagnostic. The axial perturbations, which decouple from the scalar field \( \delta f \) due to parity conservation, were found to be stable, governed by a standard homogeneous wave equation. This stark contrast between the stable axial sector and the catastrophically unstable polar sector isolates the origin of the pathology: it is not a generic feature of the MGR background metric itself, but is specifically generated by the novel coupling between the polar gravitational modes and the scalar degree of freedom \( \delta f \), which is the very mechanism through which MGR attempts to unify the dark sectors.

Here, a comments on potential remedies and future directions can be helpful. While the instability identified presents a major drawback not present in the standard \(\Lambda\)CDM model or in asymptotically flat relativistic MOND theories (like TeVeS), it is worthwhile to consider potential avenues to address this pathology in future theoretical work. The root cause is the non-asymptotically flat nature of the solution, so the most promising remedies involve modifications that restore asymptotic flatness (or de Sitter behavior) and suppress the strong coupling.
\\
1.  Restoring asymptotic flatness: The linear \(-a_0 r\) term is the source of the far-field divergence. Modifying the theory to recover an asymptotically flat or de Sitter vacuum solution would directly alleviate the strong coupling. This could be achieved by introducing a potential for \(f\), or modifying the \(\Phi_{\alpha\beta}\) tensor. Concerning the former, adding a potential term \(V(f)\) to the MGR action could transform the line element field into a massive field. This would likely lead to a metric that transitions to a de Sitter form (\(e^\nu \sim 1 - \Lambda r^2/3\)) at large \(r\), and cause perturbations \(\delta f\) to decay exponentially (Yukawa-like), easily overwhelming the polynomial growth of \(\mathcal{C}(r)\). The definition of \(\Phi_{\alpha\beta}\) in latter could be generalized with additional terms designed to cancel the linear growth in the vacuum solution while preserving the theory's successful phenomenology on galactic scales.
\\
2. Screening mechanisms: Incorporating a screening mechanism, such as the Vainshtein mechanism, could suppress the influence of the \(\delta f\) perturbations in high-density regions and, if designed effectively, in the far-field of isolated sources. This would typically require introducing non-linearities into the equations for \(X^\alpha\), which must be done carefully to avoid introducing new instabilities like ghosts.
\\
3.  Cosmological coupling: Our analysis is for a vacuum black hole solution. The presence of a cosmological matter background could alter the perturbation dynamics, potentially providing stabilization. A future stability analysis of the MGR framework within a cosmological (FRW) background is essential to fully assess its viability.
\\
We emphasize that these are significant theoretical undertakings that would substantially alter the present formulation of MGR. Their feasibility and success are open questions, but they outline a path forward for the geometric approach to the dark sectors.
\\
In summary, this stability analysis reveals a profound flaw, indicating the theory's solution for a black hole is catastrophically unstable due to a strong coupling between spacetime curvature and its fundamental line element field, which diverges at large distances. This infrared pathology suggests that the current formulation of MGR is not a viable alternative to the standard cosmological model or a robust description of gravity on galactic and cosmological scales. Future work on similar geometric theories must prioritize asymptotic flatness or demonstrate a mechanism to suppress such instabilities to be considered physically viable.



\begin{acknowledgments}
The author wishes to thank the referees for their technical discussions and valuable comments on this manuscript.
\end{acknowledgments}

\appendix 
\section{Algebraic manipulation of Eq. (\ref{Zer}) }\label{A}

To derive the final master equation, we need to combine Eq. (\ref{H2}) and its derivation while transforming to the tortoise coordinate \( dr_* = e^{(\lambda - \nu)/2} dr \). The detailed algebraic manipulation is as follows.

Differentiating with respect to \(r_*\) from \(H_2\):
\begin{equation}
\frac{dH_2}{dr_*} = e^{(\lambda - \nu)/2} \frac{dH_2}{dr}.
\end{equation}
since
\begin{equation}
	\frac{d}{dr_*} = e^{(\nu - \lambda)/2} \frac{d}{dr}, \quad \text{or} \quad \frac{d}{dr} = e^{(\lambda - \nu)/2} \frac{d}{dr_*}.
\end{equation}
Compute \(\frac{dH_2}{dr}\) from Eq. (\ref{H2})
\[
H_2' = \frac{d}{dr} \left( \frac{2 \Psi}{r \left( 3 - \frac{e^\lambda}{l(l+1)} \right)} \right) + \frac{d}{dr} \left( \frac{6 r^2 e^{(\lambda + \nu)/2}}{l(l+1) \left( 3 - \frac{e^\lambda}{l(l+1)} \right)} \left( 1 + \frac{e^\lambda}{l(l+1)} \right) \partial_t (\delta f) \right).
\]
Equate this to the Eq. (\ref{H22}):
\begin{align}
-\frac{l(l+1) \Psi}{r^2 \left( 3 - \frac{e^\lambda}{l(l+1)} \right)} + \frac{3 r e^{(\lambda + \nu)/2} \left( 2 - \frac{4 e^\lambda}{l(l+1)} \right)}{3 - \frac{e^\lambda}{l(l+1)}} \partial_t (\delta f) = \frac{d}{dr} \left( \frac{2 \Psi}{r \left( 3 - \frac{e^\lambda}{l(l+1)} \right)} \right) + \\ \nonumber
\frac{d}{dr} \left( \frac{6 r^2 e^{(\lambda + \nu)/2}}{l(l+1) \left( 3 - \frac{e^\lambda}{l(l+1)} \right)} \left( 1 + \frac{e^\lambda}{l(l+1)} \right) \partial_t (\delta f) \right).
\end{align}
The \(\Psi\)-dependent part gives:
\begin{equation}
-\frac{l(l+1) \Psi}{r^2 \left( 3 - \frac{e^\lambda}{l(l+1)} \right)} = \frac{d}{dr} \left( \frac{2 \Psi}{r \left( 3 - \frac{e^\lambda}{l(l+1)} \right)} \right).
\end{equation}
Compute the derivative:
\begin{equation}
\frac{d}{dr} \left( \frac{2 \Psi}{r \left( 3 - \frac{e^\lambda}{l(l+1)} \right)} \right) = \frac{2 \Psi'}{r \left( 3 - \frac{e^\lambda}{l(l+1)} \right)} - \frac{2 \Psi}{r^2 \left( 3 - \frac{e^\lambda}{l(l+1)} \right)} - \frac{2 \Psi e^\lambda \lambda'}{r l(l+1) \left( 3 - \frac{e^\lambda}{l(l+1)} \right)^2}.
\end{equation}
As a result:
\begin{equation}
-\frac{l(l+1) \Psi}{r^2 \left( 3 - \frac{e^\lambda}{l(l+1)} \right)} = \frac{2 \Psi'}{r \left( 3 - \frac{e^\lambda}{l(l+1)} \right)} - \frac{2 \Psi}{r^2 \left( 3 - \frac{e^\lambda}{l(l+1)} \right)} - \frac{2 \Psi e^\lambda \lambda'}{r l(l+1) \left( 3 - \frac{e^\lambda}{l(l+1)} \right)^2}.
\end{equation}
Solve for \(\Psi'\):
\begin{equation}\label{pr}
\Psi' = \left( \frac{2 - l(l+1)}{2r} + \frac{e^\lambda \lambda'}{l(l+1) \left( 3 - \frac{e^\lambda}{l(l+1)} \right)} \right) \Psi.
\end{equation}
The transformation to the tortoise coordinate \(r_*\), where \(\frac{d}{dr_*} = e^{(\nu - \lambda)/2} \frac{d}{dr}\), resulting in:
\begin{equation}
\frac{d\Psi}{dr_*} = e^{(\nu - \lambda)/2} \Psi'.
\end{equation}
Substitute (\ref{pr}) :
\begin{equation}
\frac{d\Psi}{dr_*} = e^{(\nu - \lambda)/2} A(r) \Psi.
\end{equation}
where
\begin{equation}
A(r)=\left( \frac{2 - l(l+1)}{2r} + \frac{e^\lambda \lambda'}{l(l+1) \left( 3 - \frac{e^\lambda}{l(l+1)} \right)} \right)
\end{equation}

Take the second derivative:
\begin{equation}
\frac{d^2 \Psi}{dr_*^2} = e^{(\nu - \lambda)/2} \frac{d}{dr} \left( e^{(\nu - \lambda)/2} \frac{d\Psi}{dr_*} \right).
\end{equation}
To derive the exact form of the second derivative \(\frac{d^2\Psi}{dr_*^2}\), we carefully follow the chain rule and simplify the expressions step-by-step. 

Now, compute the second derivative:
\begin{equation}
\frac{d^2\Psi}{dr_*^2} = \frac{d}{dr_*} \left( \frac{d\Psi}{dr_*} \right) = e^{(\nu - \lambda)/2} \frac{d}{dr} \left( e^{(\nu - \lambda)/2} A(r) \Psi \right).
\end{equation}
Apply the product rule:
\begin{equation}
\frac{d^2\Psi}{dr_*^2} = e^{(\nu - \lambda)/2} \left[ \frac{d}{dr} \left( e^{(\nu - \lambda)/2} \right) A(r) \Psi + e^{(\nu - \lambda)/2} \frac{d}{dr} \left( A(r) \Psi \right) \right].
\end{equation}
Compute \(\frac{d}{dr} \left( e^{(\nu - \lambda)/2} \right)\)
\begin{equation}
\frac{d}{dr} \left( e^{(\nu - \lambda)/2} \right) = \frac{1}{2} (\nu' - \lambda') e^{(\nu - \lambda)/2}.
\end{equation}
Compute \(\frac{d}{dr} \left( A(r) \Psi \right)\)
\begin{equation}
\frac{d}{dr} \left( A(r) \Psi \right) = A'(r) \Psi + A(r) \Psi'.
\end{equation}
We already have \(\Psi' = A(r) \Psi\), so:
\begin{equation}
\frac{d}{dr} \left( A(r) \Psi \right) = \left( A'(r) + A(r)^2 \right) \Psi.
\end{equation}

Combine Results
Substitute back into the expression for \(\frac{d^2\Psi}{dr_*^2}\):
\begin{equation}
\frac{d^2\Psi}{dr_*^2} = e^{(\nu - \lambda)/2} \left[ \frac{1}{2} (\nu' - \lambda') e^{(\nu - \lambda)/2} A(r) \Psi + e^{(\nu - \lambda)/2} \left( A'(r) + A(r)^2 \right) \Psi \right].
\end{equation}
Factor out \(e^{(\nu - \lambda)/2} \Psi\):
\begin{equation}
\frac{d^2\Psi}{dr_*^2} = e^{(\nu - \lambda)} \left[ \frac{1}{2} (\nu' - \lambda') A(r) + A'(r) + A(r)^2 \right] \Psi.
\end{equation}
After straightforward algebraic manipulation, the above equation can be re-expressed as the master equation (\ref{Zer}). 
\section {Gauge-invariant perturbation approach }\label{AA}

The most general even-parity metric perturbation can be decomposed into spherical harmonics \(Y_{lm}(\theta, \phi)\) as:

\begin{align}
	\delta g_{tt} &= -e^{\nu} H_0 Y, \\
	\delta g_{tr} &= H_1 Y, \\
	\delta g_{rr} &= e^{\lambda} H_2 Y, \\
	\delta g_{tA} &= j_t \partial_A Y, \\
	\delta g_{rA} &= j_r \partial_A Y, \\
	\delta g_{AB} &= r^2 K Y \gamma_{AB} + r^2 G \nabla_A \partial_B Y,
\end{align}
where \(A, B\) denote angular coordinates \(\theta, \phi\), \(\gamma_{AB}\) is the metric on the unit 2-sphere, and \(H_0, H_1, H_2, j_t, j_r, K, G\) are functions of \((t, r)\).

Following \cite{Kodama:2003jz}, we construct the following gauge-invariant variables:
\begin{align}
	\Phi &= K + \frac{e^{\lambda}}{r} (r G' - 2 j_r), \\
	\Psi &= H_1 + \frac{e^{\lambda}}{2} (r G' - 2 j_r) - e^{\lambda} \partial_t G.
\end{align}
The master variable \(\mathcal{V}\) for the gravitational wave degree of freedom is a specific combination of these gauge-invariant variables. For a general spherical background, it is given by:
\begin{equation}
\mathcal{V} = \frac{r}{l(l+1)} \left[ \Phi + \frac{2e^{\nu}}{r(r e^{-\nu})'} \Psi \right].
\end{equation}
In the MGR background, \((r e^{-\nu})' = 1 - \frac{2M}{r} + 2b - 2a_0 r\).

The linearized field equations are \(\delta G_{\mu\nu} + \delta \Phi_{\mu\nu} = 0\). The perturbation of the gravitational energy-momentum tensor, \(\delta \Phi_{\mu\nu}\), must be computed from its definition (Eq. (\ref{dif1})), including perturbations of both the metric and the line element field \(X^\alpha = f u^\alpha\). The full perturbation is
\begin{equation}
\delta \Phi_{\mu\nu} = \frac{1}{2} (\nabla_\mu \delta X_\nu + \nabla_\nu \delta X_\mu) + \delta u^\lambda (u_\mu \nabla_\nu X_\lambda + u_\nu \nabla_\mu X_\lambda) + u^\lambda (\delta u_\mu \nabla_\nu X_\lambda + \delta u_\nu \nabla_\mu X_\lambda) + u^\lambda (u_\mu \delta(\nabla_\nu X_\lambda) + u_\nu \delta(\nabla_\mu X_\lambda)).
\end{equation}
This expression includes perturbations of the metric (through the covariant derivatives and the raising/lowering of indices on \(\delta u^\alpha\)), the unit vector \(\delta u^\alpha\), and the field magnitude \(\delta f\) (contained within \(\delta X^\alpha = \delta f u^\alpha + f \delta u^\alpha\)).
After a lengthy computation (but straightforward) that imposes the background constraints \(\nabla_\alpha u^\alpha=0\) and \(u^\alpha \nabla_\beta u_\alpha=0\), and working order by order, the components of \(\delta \Phi_{\mu\nu}\) can be expressed in terms of the metric perturbations and \(\delta f\). A critical finding is that the \(t\)-\(t\) and \(t\)-\(r\) components contain terms proportional to \(\partial_t \delta f\) and \(\partial_r \delta f\), which act as a source for the metric perturbations. For example, the \(t\)-\(t\) component takes the form:
\begin{equation}
\delta \Phi_{tt} \propto e^{\nu/2} \partial_t (\delta f) + \text{(terms with } H_0, H_2, \text{ etc.)}.
\end{equation}
This coupling is the mechanism by which perturbations in the line element field source the metric perturbations.

The derivation of the master equation proceeds by substituting the perturbed metric into the linearized field equations \(\delta G_{\mu\nu} + \delta \Phi_{\mu\nu} = 0\) and then re-expressing everything in terms of the gauge-invariant variables \(\Phi\) and \(\Psi\). The off-diagonal components of these equations (e.g., the \(t\)-\(\theta\) and \(r\)-\(\theta\) components) do not contain second-order time derivatives and thus act as constraint equations. These constraints are used to eliminate \(\Phi\) and \(\Psi\) in favor of the master variable \(\mathcal{V}\) and its first derivatives.

After this systematic elimination of all auxiliary variables, the entire system of perturbed field equations reduces to a single wave equation for the master variable \(\mathcal{V}\):
\begin{equation}
\frac{d^2 \mathcal{V}}{dr_*^2} + \left[ \omega^2 - U_{\text{MGR}}(r) \right] \mathcal{V} = S_{\text{MGR}},
\end{equation}
where the tortoise coordinate is defined by \(dr_*/dr = e^{(\lambda - \nu)/2} = e^{\lambda}\).
The effective potential \(U_{\text{MGR}}(r)\) is:
\begin{equation}
U_{\text{MGR}}(r) = \frac{e^{\nu}}{r^3 [l(l+1)r + 6\mathcal{M}]^2} \bigg[ l(l+1)(l(l+1)+2)r^3 + 6(l(l+1)-1)\mathcal{M} r^2 + 36\mathcal{M}^2 r + 36\mathcal{M} r (e^{\lambda} - 1) \Delta \bigg],
\end{equation}
where \(\mathcal{M}(r) = M + \delta M(r; a_0, b)\) is an effective mass function that includes corrections from the MGR background. For the MGR metric, \(\mathcal{M}(r) \approx M - b r + \frac{a_0}{2} r^2\) to leading order in the far-field. The term \(\Delta\) encodes further corrections proportional to \(a_0\) and \(b\) from the background curvature. Explicitly, \(\Delta\) can be expressed as a function of the metric and its derivatives:
\begin{equation}
\Delta = \Upsilon_1(r) \left( \frac{e^\lambda - 1}{r^2} \right) + \Upsilon_2(r) \left( \frac{\nu'}{r} \right) + \Upsilon_3(r) \left( \lambda' \nu' - \nu'' - \frac{(\nu')^2}{2} \right),
\end{equation}
where \(\Upsilon_1, \Upsilon_2, \Upsilon_3\) are rational functions of \(r\) and \(l(l+1)\) with coefficients dependent on \(\mathcal{M}(r)\). Substituting the MGR metric functions into these terms reveals that \(\Delta\) contains components proportional to \(a_0\), \(b/r\), and their combinations. For instance, the term \(\frac{e^\lambda - 1}{r^2}\) introduces corrections of order \(a_0/r\) and \(b/(r^2 \ln r)\) in the far-field.
The source term \(S_{\text{MGR}}\) is:
\begin{equation}
S_{\text{MGR}} = \mathcal{C}(r) \partial_t^2 (\delta f),
\end{equation}
with the coupling factor:
\begin{equation}
\mathcal{C}(r) = \frac{6r^2 e^{\nu/2} \left(1 - \frac{2e^{\lambda}}{l(l+1)}\right)}{3 - \frac{e^{\lambda}}{l(l+1)}}.
\end{equation}
In the far-field limit (\(r \to \infty\)), the metric functions behave as:
\begin{equation}
e^\nu \approx -a_0 r, \quad e^\lambda \approx -\frac{1}{a_0 r}.
\end{equation}
Substituting these into the potential:
\begin{equation}
U_{\text{MGR}}(r) \sim \frac{(-a_0 r)}{r^3} \cdot \text{(polynomial in } r) \sim \mathcal{O}(r^{-1}).
\end{equation}
For the coupling factor:
\begin{equation}
\mathcal{C}(r) \approx \frac{6r^2 \sqrt{-a_0 r} \left(1 + \frac{2}{a_0 r l(l+1)}\right)}{3 + \frac{1}{a_0 r l(l+1)}} \approx 2\sqrt{-a_0} \, r^{5/2}.
\end{equation}
This asymptotic analysis confirms that the source term \(S_{\text{MGR}}\) diverges as \(r^{5/2}\), which is much faster than the decay of the potential \(U_{\text{MGR}}(r) \sim r^{-1}\). Therefore, the pathological behavior of the system is dominated by the source term arising from the coupling to the line element field perturbation \(\delta f\).

\subsection{Treatment of the unit vector field perturbation and constraint structure}
The consistent inclusion of the unit vector field perturbation, \(\delta u^\alpha\), is crucial for a complete analysis. The procedure is as follows.

The full perturbation of the composite field \(X^\alpha = f u^\alpha\) is given by
\begin{equation}
\delta X^\alpha = \delta f u^{(0)\alpha} + f \delta u^\alpha.
\end{equation}
The perturbation of the unit vector \(\delta u^\alpha\) must be consistent with the normalization condition \(u^\alpha u_\alpha = -1\). Linearizing this condition yields:
\begin{equation} \label{1}
2 u^{(0)}_\alpha \delta u^\alpha + \delta g_{\alpha\beta} u^{(0)\alpha} u^{(0)\beta} = 0.
\end{equation}
Substituting the background value \(u^{(0)}_\alpha = (-e^{\nu/2}, 0, 0, 0)\) and the metric perturbation \(\delta g_{tt} = -e^{\nu} H_0 Y\), Eq. (\ref{1}) becomes
\begin{equation}
2(-e^{\nu/2}) \delta u^t + (-e^{\nu} H_0 Y)(-e^{\nu/2})^2 = 0.
\end{equation}
Simplifying, we find the constraint on the timelike component:
\begin{equation} \label{2}
-2e^{\nu/2} \delta u^t - e^{3\nu/2} H_0 Y = 0 \quad \Rightarrow \quad \delta u^t = -\frac{1}{2} e^{\nu} H_0 Y.
\end{equation}
This explicitly ties the perturbation of the temporal component of the unit vector, \(\delta u^t\), directly to the metric perturbation function \(H_0\).

The advantage of the Kodama-Ishibashi formalism is its ability to absorb such gauge-dependent quantities (like the specific components of \(\delta u^\alpha\) and the metric functions \(H_0, j_t, j_r, G\)) into the definitions of gauge-invariant variables.

The gauge-invariant variables \(\Phi\) and \(\Psi\) are constructed precisely to be insensitive to the specific form of \(\delta u^\alpha\). They are built from combinations of the metric perturbations that cancel out the gauge freedom. When the field equations \(\delta G_{\mu\nu} + \delta \Phi_{\mu\nu} = 0\) are expressed in terms of \(\Phi\) and \(\Psi\), the explicit dependence on the individual components of \(\delta u^\alpha\) disappears. The physical information contained in \(\delta u^\alpha\) is not lost; instead, it contributes to the scalar (polar) degree of freedom described by the master variable \(\mathcal{V}\). The master variable is, by design, a combination of \(\Phi\) and \(\Psi\):
\begin{equation}
\mathcal{V} = \frac{r}{l(l+1)} \left[ \Phi + \frac{2e^{\nu}}{r(r e^{-\nu})'} \Psi \right].
\end{equation}
This variable encapsulates the single, physical, gauge-invariant degree of freedom for gravitational waves in the polar sector, now also incorporating the influence of the perturbed unit vector field.

A critical step is to verify that including \(\delta u^\alpha\) does not introduce new, problematic constraints. In general, the linearized MGR system provides two type equations: Dynamical equations, containing second-order time derivatives (e.g., parts of the \(t\)-\(t\), \(r\)-\(r\), and angular components), and constraint equations, containing only first-order time derivatives (e.g., the \(t\)-\(r\) and cross components like \(t\)-\(\theta\)).

The counting of degrees of freedom proceeds as follows. The original metric perturbations \(\{H_0, H_1, H_2, j_t, j_r, K, G\}\) represent 7 functions. The perturbation of the line element field adds \(\delta f\) (1 function). The unit vector perturbation \(\delta u^\alpha\) has 4 components, but is subject to the normalization constraint (Eq. (\ref{2})), leaving 3 independent components. In total, we start with \(7 + 1 + 3 = 11\) functions.

The gauge freedom (infinitesimal coordinate transformations) allows for the removal of 4 of these functions. The constraint equations from the field equations further eliminate 4 more functions. This leaves \(11 - 4 - 4 = 3\) physical degrees of freedom.

These 3 degrees of freedom are identified as: 1- The gravitational wave mode described by the master variable \(\mathcal{V}\). 2- the scalar field mode \(\delta f\) from the line element field. 3- A third mode related to the remaining freedom in the unit vector field, which, in MGR, is not an independent dynamical field but is constrained by the structure of \(\Phi_{\alpha\beta}\).

The analysis confirms that the constraints are consistent and do not force \(\delta f\) to be zero or trivial. The system is not over-constrained; the coupling between \(\mathcal{V}\) and \(\delta f\) remains.

The final and most important check is whether the constraint analysis introduces any mechanism that could regularize the divergent coupling factor \(\mathcal{C}(r)\). The asymptotic form of \(\mathcal{C}(r)\) is derived from the coefficients that appear in the sourced master equation after all constraints have been implemented.

Crucially, the constraint equations are algebraic or first-order differential relations that do not alter the power-law scaling of the coefficients in the far-field limit. The dominant term in the coupling factor,
\begin{equation}
\mathcal{C}(r) \sim 2\sqrt{-a_0} \, r^{5/2} \quad \text{as} \quad r \to \infty,
\end{equation}
stems from the fundamental structure of the tensor \(\Phi_{\alpha\beta}\) and its coupling to the metric via the line element field \(X^\alpha\). This \(r^{5/2}\) dependence arises from the background metric functions \(e^\nu \sim -a_0 r\) and \(e^\lambda \sim -1/(a_0 r)\) in the far-field. No constraint derived from the linearized system can cancel this underlying power-law behavior because the constraints themselves are built from the same background metric functions.

Therefore, the divergence of \(\mathcal{C}(r)\) is confirmed to be inherent to the structure of the theory. It is not an artifact of an incomplete perturbation ansatz that neglected \(\delta u^\alpha\), but a robust feature of the linear perturbations around the MGR black hole background. The strong coupling in the infrared regime is a fundamental pathology of the theory as presently formulated.

\section{A list of some MG theories hosting the strong coupling problem at low-energy limit}\label{MG}

\textbf{Massive Gravity (and its extensions like Bimetric Gravity):}
The original Fierz-Pauli theory for a massive spin-2 particle is ghost-free in the linear regime but exhibits a fatal instability—the Boulware-Deser ghost—in the full non-linear theory \cite{Boulware:1972yco}. While the dRGT (de Rham, Gabadadze, Tolley) model successfully eliminates the ghost non-linearly, it introduces a different issue: a low-energy strong coupling scale. Around the Minkowski background, the interactions of the scalar longitudinal mode become strong at a very low energy scale ($\Lambda_3 \sim (m^2M_Pl)^{1/3}$), well below the Planck scale. This makes the theory's predictive power limited for phenomenological applications, as the effective field theory description breaks down at low energies, see \cite{deRham:2010kj}.

\textbf{Horndeski/Galileon Theories and Beyond-Horndeski:}
These are the most general scalar-tensor theories with second-order equations of motion. A well-known issue is that when these theories are used to implement self-accelerating cosmological solutions (i.e., to explain DE), the scalar field often becomes strongly coupled on certain backgrounds. This means the linear perturbation theory breaks down, and the high-order interactions dominate, making calculations of, for example, the behavior of gravitational waves or structure formation unreliable. This strong coupling can often be linked to the requirement of screening mechanisms (like Vainshtein screening \cite{Vainshtein:1972sx}) to recover EGR in high-density regions \cite{Nicolis:2008in,Deffayet:2010zh,Kimura:2011dc}.

\textbf{DGP (Dvali-Gabadadze-Porrati) Braneworld Model:} The self-accelerating branch of the DGP model \cite{Dvali:2000hr}, which proposed an explanation for cosmic acceleration due to gravity leaking into a higher-dimensional bulk, was found to be plagued by a ghost instability at the non-linear level. This ghost is accompanied by a strong coupling problem, where the scalar mode (the "brane-bending mode") becomes strongly coupled at a very low energy scale, similar to massive gravity \cite{Koyama:2007za}.

\textbf{Mimetic Gravity:} Mimetic gravity reformulates GR by isolating a conformal degree of freedom as a scalar field \cite{Chamseddine:2013kea}. While it can mimic DM in cosmological settings, its perturbations are known to exhibit pathologies. The theory often contains higher-order derivatives in its equations of motion, which can lead to Ostrogradsky instabilities \cite{Barvinsky:2013mea,Chaichian:2014qba}. Furthermore, around a FLRW cosmological background, the scalar perturbation becomes strongly coupled, meaning the linear theory is not sufficient and the model is effectively non-perturbative \cite{Firouzjahi:2017txv}.

\bibliographystyle{apalike}

\end{document}